\documentclass{aa}  
\usepackage{graphicx}
\usepackage{txfonts}
\usepackage{aalongtable}
\begin{document}
\input{psfig}
%
\newbox\grsign \setbox\grsign=\hbox{$>$} \newdimen\grdimen \grdimen=\ht\grsign
\newbox\simlessbox \newbox\simgreatbox
\setbox\simgreatbox=\hbox{\raise.5ex\hbox{$>$}\llap
     {\lower.5ex\hbox{$\sim$}}}\ht1=\grdimen\dp1=0pt
\setbox\simlessbox=\hbox{\raise.5ex\hbox{$<$}\llap
     {\lower.5ex\hbox{$\sim$}}}\ht2=\grdimen\dp2=0pt
\def\simgreat{\mathrel{\copy\simgreatbox}}
\def\simless{\mathrel{\copy\simlessbox}}
\newbox\simppropto
\setbox\simppropto=\hbox{\raise.5ex\hbox{$\sim$}\llap
     {\lower.5ex\hbox{$\propto$}}}\ht2=\grdimen\dp2=0pt
\def\simpropto{\mathrel{\copy\simppropto}}
\makeatletter
\renewcommand{\table}[1][]{\@float{table}[!htp]}
\makeatother 
\title{High-resolution abundance analysis of red giants in  the globular cluster NGC 6522 
\thanks{Observations collected at the European Southern Observatory,
Paranal, Chile (ESO), under programmes 88.D-0398A.}}


\author{
B. Barbuy\inst{1}
\and
C. Chiappini\inst{2}
\and
E. Cantelli\inst{1}
\and
E. Depagne\inst{2}
\and
M. Pignatari\inst{3}
\and
R. Hirschi\inst{4}
\and
G. Cescutti\inst{2}
\and
S. Ortolani\inst{5,6}
\and
V. Hill\inst{7}
\and
M. Zoccali\inst{8,9}
\and
D. Minniti\inst{9,10}
\and
M. Trevisan\inst{11}
\and
E. Bica\inst{12}
\and
A. G\'omez\inst{13}
}
\offprints{B. Barbuy}

\institute{
Universidade de S\~ao Paulo, IAG, Rua do Mat\~ao 1226,
Cidade Universit\'aria, S\~ao Paulo 05508-900, Brazil;
e-mail: barbuy@astro.iag.usp.br
\and
Leibniz-Institut f\"ur Astrophysik Potsdam (AIP), 
An der Sternwarte 16, 14482, Potsdam, Germany;
e-mail: cristina.chiappini@aip.de
\and 
Department of Physics, University of Basel, Klingelbergstrasse 82,
 Basel, 4056, Switzerland
\and
Astrophysics Group, Keele University, ST5 5BG, Keele, England
\and
Dipartimento di Fisica e Astronomia, Universit\`a di Padova, I-35122 Padova,
 Italy
\and
INAF-Osservatorio Astronomico di Padova, Vicolo dell'Osservatorio 5,
I-35122 Padova, Italy
\and
Laboratoire Lagrange (UMR7293), Universit\'e de Nice Sophia Antipolis,
 CNRS, Observatoire de la C\^ote d’Azur, CS 34229, F-06304 Nice cedex 4, France
\and
Universidad Catolica de Chile, Instituto de Astrofisica,
Casilla 306, Santiago 22, Chile
\and
Millenium Institute of Astrophysics, Av. Vicu\~na Mackenna 4860,
Macul, Santiago, Chile
\and
Departamento de Ciencias Fisicas, Universidad Andres Bello, Republica 220, Santiago, Chile
\and
Instituto Nacional de Pesquisas Espaciais, Av. dos Astronautas 1758, S\~ao Jos\'e dos Campos 12227-010, Brazil
\and
Universidade Federal do Rio Grande do Sul, Departamento de Astronomia,
CP 15051, Porto Alegre 91501-970, Brazil
\and
Observatoire de Paris-Meudon, GEPI, 92195 Meudon Cedex, France
}
 
   \date{accepted in 31/07/2014 for Astronomy \& Astrophysics}

 \abstract
   {
The [Sr/Ba] and [Y/Ba] scatter observed in some galactic halo
stars that are very metal-poor stars and in a few individual stars of 
 the oldest known Milky Way globular cluster NGC 6522,
have been interpreted as evidence of early enrichment  by massive 
fast-rotating stars (spinstars).  
Because NGC 6522 is a bulge globular cluster, the suggestion was that not only the very-metal poor halo stars, but also bulge stars at [Fe/H] $\sim -$1 could be used as probes of the stellar nucleosynthesis signatures from the earlier generations of massive stars, but at much higher metallicity.  
For the bulge the suggestions were based on early spectra available for stars in NGC 6522,  with a medium resolution of R$\sim$22,000 and 
a moderate signal-to-noise ratio.}
   {The  main purpose  of this  study is to re-analyse the NGC 6522 stars
previously reported using new high-resolution (R$\sim$45,000) and 
high signal-to-noise spectra (S/N$>$100).
 We aim at re-deriving their stellar parameters and elemental  ratios,
in particular the abundances of the neutron-capture 
s-process-dominated elements such as Sr, Y, Zr, La, and Ba, and of the
 r-element Eu.}
   {High-resolution spectra of four giants belonging to
  the bulge globular cluster NGC 6522
 were obtained at the 8m VLT UT2-Kueyen  telescope
 with the UVES spectrograph in FLAMES-UVES
configuration. The spectroscopic parameters were derived based on 
the excitation and ionization equilibrium of \ion{Fe}{I} and \ion{Fe}{II}.}
   {Our analysis confirms a metallicity  [Fe/H] = $-0.95\pm0.15$ for NGC 6522,
and the overabundance of the studied stars in Eu (with +~0.2 $<$ [Eu/Fe] $<$ +~0.4) and alpha-elements O and Mg. 
The neutron-capture s-element-dominated Sr, Y, Zr, Ba, La now show
less pronounced variations from star to star. Enhancements are in the
 range 0.0 $<$ [Sr/Fe] $<$ +0.4, 
+0.23 $<$ [Y/Fe] $<$ +0.43, 0.0 $<$ [Zr/Fe] $<$ +0.4, 0.0 $<$ [La/Fe] $<$ +0.35,
and  0.05 $<$ [Ba/Fe] $<$ +0.55.}
{   }
   
\titlerunning{Abundance analysis of four giants in NGC 6522}
\authorrunning{B. Barbuy et al.}
   \maketitle
%

\section{Introduction}

The heavy-element abundances in very old stars have been interpreted
by Truran (1981) to correspond to r-process products, because the
 s-pro\-cess en\-rich\-ment can\-not pro\-ceed promp\-tly in the early Galaxy.
 On the other
hand, it is well\--known that ro\-ta\-tion in mas\-sive stars can explain high
primary nitrogen abundances in low-metallicity stars, because of internal
mixing induced by rotation (e.g. Barbuy 1983; Chiappini et al. 2006).
It has now been shown that s-elements can also be produced
 in fast-rotating massive stars, or spinstars
(Pignatari et al. 2008; Frischknecht et al. 2012). 
Chiappini (2013) described
the impact of spinstars on the chemical enrichment of the early Universe,
 and how some of the very metal-poor halo data can be better matched
 when the spinstar's contribution is taken into account.

Chiappini et al. (2011, hereafter C11) reported the first s-process
 detailed nucleosynthesis calculations made by 
Frischknecht and collaborators
 for a fast-rotating massive star of  40 
solar masses, with a metallicity [Fe/H] = $-$3.8  and rotational velocity 
V$_{rot}$ = 500 km~s$^{-1}$. It was shown that s-process elements in this star 
were boosted by up to four orders of magnitude with respect to a non-rotating star of same mass and metallicity  (see their Fig. 2). An extended grid of spinstar models was later computed and published in Frischknecht et al. (2012).
 In particular, these calculations indicate an enhancement of the heavy 
elements Sr, Y, La, and Ba that are measurable in stellar spectra.

The yields of the first very metal-poor supernovae in the Galactic halo
can be studied in the most metal-poor stars (e.g. Cayrel et al. 2004),
whereas in the Galactic bulge a star formation rate enhanced by a factor
10 relative to the halo (e.g. Ballero et al. 2007) 
resulted in the oldest bulge stars having [Fe/H]$\sim-$1.0 (Chiappini et al. 2014, in prep.).
Evidence for the oldest bulge stellar populations having [Fe/H]$\sim-$1.0
was reported in Lee (1992) and D\'ek\'any et al. (2013), in studies of bulge 
RR Lyrae with this metallicity. Also, a
 significant number of very old globular clusters with
 this metallicity was also found in the inner bulge 
(Minniti 1995; Rich et al. 1998; Barbuy et al. 2009, hereafter B09).
 
In C11 the stellar yields of the 40 solar mass spinstar were compared
 with the [Y/Ba] and [Sr/Ba] ratios observed in very metal-poor halo stars 
and with the abundances derived for 
red giant stars of the bulge metal-poor globular cluster  NGC 6522. 
These abundances were first reported by Barbuy et al. (2009, hereater B09). 
C11 suggested that bulge stars with a metallicity of
 [Fe/H]$\sim$$-$1 might also hold information on the nature of the first generations of stars, as has been known to be the case for halo stars with  metallicities below [Fe/H] $\simeq -$3 (e.g. Truran 1981; Cayrel et al. 2004). Although there are bulge stars with lower metallicities, the bulge metallicity distribution shows a sharp cutoff around [Fe/H] = $-$1 (e.g. Ness et al. 2014; Zoccali et al. 2008). Despite uncertainties in the NGC 6522 data, the eight stars measured showed a scatter in the [Y/Ba] ratio similar to that observed in 
very metal-poor halo stars (with [Fe/H] $<-$3). 

 In C11 it was then suggested that the highest values of the [Y/Ba] observed in the halo (around [Fe/H] = $-$ 3) and bulge (around [Fe/H] = $-$1) was caused by the s-process contribution of spinstars. Indeed, according to spinstar models, their production would be strongly dependent on mass and 
rotational-velocity, which would explain the observed scatter in the ratio of two chemical elements dominated by the s-process nucleosynthesis of two different s-process peaks. Although this explanation seemed unique for three
 stars of NGC 6522, other alternative explanations for stars with lower [Y/Ba] ratio at metallicities around [Fe/H] = $-$1 cannot be discarded, 
such as the AGB mass-transfer contribution (Bisterzo et al. 2011). Moreover, it was also clear that the spinstar contribution is probably complementary to that coming from r-process nucleosynthetic sites (e.g. Goriely et al. 2013;  Nakamura et al. 2013; Wanajo 2013;  Qian 2012;
 Winteler et al. 2012; Arcones \& Mar\'{\i}nez-Pinedo 2011).

More recently, it was possible to quantitatively test these ideas
for halo stars thanks to the more complete grid of spinstar
 stellar models provided in Frischknecht et al. (2012). With this grid 
we computed inhomogeneous chemical evolution models for the halo 
(Cescutti et al. 2013; Cescutti \& Chiappini 2014) and showed
 that indeed the observed scatter in abundance ratios of two
predominantly s-process elements in stars with metallicities between
 $-$4 $<$ [Fe/H] $< -$3 can be well explained 
if the s-process contribution of spinstars in the
early chemical enrichment of the Universe is considered.
 Aoki et al. (2013) proposed an alternative scenario for the scatter in [Sr/Ba].
 Their suggestion is that the explosion of core
 collapse SNe could produce r-process,
 but in some cases the r-process-rich material is not ejected 
because of the subsequent
 collapse of the proton-neutron star to a black hole. On the other hand,
  there is currently no clear evidence from 
Galactic chemical evolution simulations that
 explosive primary processes alone can
 explain the observed distribution of [Sr/Ba] with respect to the [Fe/H]
 in the Galactic halo, with an observed peak of at about [Fe/H] $\sim -$3,
 while this is the case for the s-process from spinstars (Cescutti 
\& Chiappini 2014). 
This result is also confirmed taking into account a recent
 re-analysis of Sr observations
 in a large sample of metal-poor stars (Hansen et al. 2013).
 This peculiar feature of the s-process from spinstars is due to the intrinsic 
nature of the s-process in spinstars, which is not primary even though
 the main neutron source does not depend on the initial
 metallicity of the star (Pignatari et al. 2008; Frischknecht et al. 2012).
 For the bulge (Chiappini et al. 2014, in prep.), we instead find that the scatter in these abundance ratios is smaler, which is a consequence of the
 higher star formation rate. Interestingly, these new models cannot explain the large overabundances reported in B09 and C11. As  we will
 show here, the new high-quality spectra analysed suggest
 lower [Sr/Ba] and [Y/Ba] than what was reported before based on
 FLAMES-GIRAFFE lower resolution spectra, and hence agree well
 with our model predictions.


  
We present results from new spectra obtained with
the FLAMES-UVES spectrograph at the VLT,  at a resolution R$\sim$45,000, 
and S/N$\sim$ 90-130 in the red portion
(580-680nm), and 70-130 in the blue portion (480-580nm),
 to verify the previous abundance results.
 A detailed abundance analysis  of four stars  in
NGC 6522 based on these higher resolution spectra
is carried out using MARCS model atmospheres (Gustafsson et al. 2008).

   The observations are described  in Sect. 2.
  Photometric stellar parameters effective temperature and  gravity  
are   derived in  Sect. 3.  Spectroscopic parameters are
derived in  Section  4, and   abundance ratios are   computed in  Sect.
5. A discussion is presented in Sect. 6 and conclusions are drawn in Sect. 7.

\section{Observations} 

The spectra of individual stars of NGC 6522
 were obtained at the VLT using the UVES spectrograph
(Dekker et al. 2000) in FLAMES-UVES mode. 
 The red chip (5800-6800 {\rm \AA}) has the
ESO CCD \# 20, an MIT backside illuminated, of 4096x2048 pixels, and pixel
size  15x15$\mu$m.
The blue chip (4800-5800 {\rm \AA}) uses the ESO Marlene EEV
CCD\#44, backside illuminated, of 4102x2048 pixels, 
and the pixel size is  15x15$\mu$m. 
The UVES standard setup 580 yields a resolution R $\sim$ 45,000 for a
slit width of 1 arcsec.
 The pixel scale is 0.0147 {\rm \AA}/pix, 
with $\sim$7.5 pixels  per resolution element at 6000 {\rm \AA}.


 The data were reduced using the 
UVES pipeline, within the ESO/Reflex software (Ballester et al. 2000;
Modigliani et al. 2004). 
The log of the 2011-2012 observations is given in  Table \ref{logobs}. 
 The   spectra   were  flatfielded,  optimally extracted,   and
wavelength calibrated    with the  FLAMES-UVES pipeline.
Spectra  extracted  from different frames were  then  co-added, taking into 
account the radial velocities reported in Table \ref{vr}.

Previous observations by Zoccali et al. (2008) and B09
 were taken in the wavelength range
$\lambda\lambda$ 6100-6860 {\rm   \AA}, using 
 the GIRAFFE setups HR13 ($\lambda\lambda$ 6120-6402 {\rm \AA}), 
 HR14 ($\lambda\lambda$ 6381-6620 {\rm \AA}),
 and HR15 ($\lambda\lambda$ 6605-6859 {\rm \AA}) at a resolution
R=22,000 and S/N = 80-100.

The present UVES observations centred on 5800 {\rm \AA} yield
a spectral coverage of 4800 $<$ $\lambda$ $<$ 6800 {\rm \AA},
with a gap at 5708-5825 {\rm \AA}.
Five of the eight stars analysed in B09 and
C11 were observed. The spectrum of B-108,
however, appears to correspond to a much hotter star than in the
early data (see detailed discussion in Sect. \ref{b108})
                                 
Table~\ref{starmag} gives the selected stars, 
their OGLE and 2MASS designations, 
 coordinates, and the $V$, $I$ magnitudes 
 from the OGLE-II catalogue (Udalski et al. 2002)
 together with the 2MASS JHK$_s$ (Skrutskie et al. 2006) 
and VVV JHK$_s$ magnitudes (Saito et al. 2012).

\begin{table}
\caption[1]{Log of the spectroscopic observations carried out on October 8, 
2011 (Julian Date 2455842), and 
2012 March 6-7 and 25 (Julian Dates 2455992, 2455993, 2456011).
 The quoted seeing is the mean value along the exposures on the detector.}
\begin{flushleft}
\begin{tabular}{lllllllccccccc}
\noalign{\smallskip}
\hline
\noalign{\smallskip}
\hline
\noalign{\smallskip}
Date & UT & exp   & Airmass  & Seeing & \\
\noalign{\smallskip}
 &  &  (s) &   & ($''$) &  \\
\noalign{\smallskip}
\noalign{\smallskip}
\hline
\noalign{\vskip 0.2cm}
 08.10.11 & 00:45:54.5 & 2750  & 1.328-1.595 & 0.82$''$  &\\
 08.10.11 & 01:34:37.2 & 2750  & 1.611-2.096 & 1.29$''$  &\\
 06.03.12 & 07:38:32.6 & 2750  & 1.743-1.414 & 1.15$''$  &\\
 06.03.12 & 08:28:44.5 & 2750  & 1.343-1.176 & 0.93$''$  &\\
 07.03.12 & 07:47:56.9 & 2750  & 1.630-1.348 & 0.81$''$  &\\
 07.03.12 & 08:39:16.9 & 2750  & 1.372-1.167 & 0.73$''$  &\\
 25.03.12 & 08:31:47.0 & 2750  & 1.127-1.048 & 0.64$''$  &\\
\noalign{\smallskip}
\hline
\end{tabular}
\end{flushleft}
\label{logobs}
\end{table}

\begin{table}
\caption[2]{Radial velocities of the UVES sample stars, in each of the seven
exposure runs with corresponding heliocentric radial velocities and mean
heliocentric radial velocity.}
\small
\begin{flushleft}
\begin{tabular}{l@{}c@{}c@{}c@{}c@{}c@{}c@{}c@{}}
\noalign{\smallskip}
\hline
\noalign{\smallskip}
\hline
\noalign{\smallskip}
Target & \phantom{-}${\rm v_r^{obs}}$ & \phantom{-}\phantom{-}${\rm v_r^{hel.}}$ & & \phantom{-}Target & \phantom{-}${\rm v_r^{obs}}$ & 
\phantom{-}\phantom{-}${\rm v_r^{hel.}}$ \\
\noalign{\smallskip}
&${\rm km~s^{-1}}$ &${\rm km~s^{-1}}$ & & &${\rm km~s^{-1}}$ &${\rm km~s^{-1}}$ \\
\noalign{\smallskip}
\noalign{\smallskip}
\hline
\noalign{\vskip 0.2cm}
B-107 1 & \phantom{-}+21.705 & \phantom{-}$-$7.225 & & \phantom{-}B-128 1 & \phantom{-}+14.745 & \phantom{-}$-$14.186 & \\
B-107 2 & \phantom{-}+21.594 &  \phantom{-}$-$7.377  & & \phantom{-}B-128 2 & \phantom{-}+14.559 & \phantom{-}$-$14.401 &  \\
B-107 3 & \phantom{-}$-$37.038 & \phantom{-}$-$8.008 & & \phantom{-}B-128 3 & \phantom{-}$-$43.743 & \phantom{-}$-$14.713 & \\
B-107 4 & \phantom{-}$-$36.955 & \phantom{-}$-$7.965 & & \phantom{-}B-128 4 & \phantom{-}$-$43.725 & \phantom{-}$-$14.735 &  \\
B-107 5 & \phantom{-}$-$36.715 & \phantom{-}$-$7.575 & & \phantom{-}B-128 5 & \phantom{-}$-$43.990 & \phantom{-}$-$14.850 & \\
B-107 6 & \phantom{-}$-$36.705 & \phantom{-}$-$7.605 & & \phantom{-}B-128 6 & \phantom{-}$-$43.928 & \phantom{-}$-$14.828 & \\
B-107 7 & \phantom{-}$-$37.440 & \phantom{-}$-$7.630 & & \phantom{-}B-128 7 & \phantom{-}$-$44.657 & \phantom{-}$-$14.847 & \\
\hbox{Mean} & --- & \phantom{-}$-$7.626 & &
 \hbox{Mean} & --- & \phantom{-}$-$14.651 & \\
\noalign{\smallskip}
\hline
\noalign{\vskip 0.2cm}
B-122 1 & \phantom{-}+11.333 & \phantom{-}$-$17.597 & & \phantom{-}B-130 1 & \phantom{-}+12.351 & \phantom{-}$-$16.579 & \\
B-122 2 & \phantom{-}+11.156 & \phantom{-}$-$17.815 & & \phantom{-}B-130 2 & \phantom{-}+12.105 & \phantom{-}$-$16.865 & \\
B-122 3 & \phantom{-}$-$47.488 & \phantom{-}$-$18.458 & & \phantom{-}B-130 3 & \phantom{-}$-$46.004 & \phantom{-}$-$16.974 & \\
B-122 4 & \phantom{-}$-$47.500 & \phantom{-}$-$18.510 & & \phantom{-}B-130 4 & \phantom{-}$-$46.049 & \phantom{-}$-$17.059 &  \\
B-122 5 & \phantom{-}$-$47.059 & \phantom{-}$-$17.919 & & \phantom{-}B-130 5 & \phantom{-}$-$45.779 & \phantom{-}$-$16.639 & \\
B-122 6 & \phantom{-}$-$47.133 & \phantom{-}$-$18.032 & & \phantom{-}B-130 6 & \phantom{-}$-$45.849 & \phantom{-}$-$16.749 & \\
B-122 7 & \phantom{-}$-$47.778 & \phantom{-}$-$17.968 & & \phantom{-}B-130 7 & \phantom{-}$-$46.601 & \phantom{-}$-$16.791 & \\
\hbox{Mean} & & \phantom{-}$-$18.043 & &\hbox{Mean} & --- & \phantom{-}$-$16.808 & \\
\noalign{\smallskip}
\hline
\end{tabular}
\end{flushleft}
\label{vr}
\end{table}

\section{Radial velocities}

 In Table \ref{vr} we report the radial velocities measured with IRAF/FXCOR
for each of the seven runs, together with their mean values. A mean heliocentric
radial velocity of v$^{\rm hel}_{\rm r}$ = $-$14.3$\pm$0.5 km~s$^{\rm {-1}}$
is obtained.

 The radial velocities reported in B09 were revised. It was verified that
the values given in B09 for setups HR14 and HR15 do not correspond
to their radial velocities, given that these spectra had been shifted in
wavelength, when they were reduced as presented in Zoccali et al. (2008).
The observed radial velocities v$_{\rm r}$ were  measured in each
 of the setups, and the HR14 and HR15 spectra were shifted to the 
same v$_{\rm r}$ zero point as that of HR13 (by means of an empirical 
differential heliocentric correction measured as the mean
 shift in radial velocity of all the stars, between a given
 setup and HR13). An average was taken
    as if all the spectra were taken on the same day as HR13. Then
    the heliocentric correction to  v$_{\rm r}$ was calculated for
    the HR13 setups  and was applied to all spectra.
Therefore the vr values from HR14 and HR15 given in B09
 should be disconsidered, as well as the mean value of
 v$^{\rm hel}_{\rm r}$ = -24.9$\pm$0.7 km~s$^{\rm {-1}}$ given in B09.

The HR13 setup values given in B09 are the correct ones. Therefore
the radial velocities are  v$^{\rm hel}_{\rm r}$ =
 $-$6.59, $-$15.86, $-$12.28, and $-$15.67 km~s$^{\rm {-1}}$ for stars
B-107, B-122, B128, and B-130 respectively. This results in a mean value
of  v$^{\rm hel}_{\rm r}$ = $-$12.6$\pm$0.7 km~s$^{\rm {-1}}$ for
the GIRAFFE spectra, program 071.D-0617(A), that were studied in B09.


The present mean heliocentric radial velocity v$^{\rm hel}_{\rm r}$ =
 $-$14.3$\pm$0.5 km~s$^{\rm {-1}}$ found for NGC 6522 from the UVES
spectra agrees well with
the value v$^{\rm hel}_{\rm r}$ = $-$12.6$\pm$0.7 km~s$^{\rm {-1}}$
from the GIRAFFE data.
It also agrees well with
v$^{\rm hel}_{\rm r}$ = $-$18.3$\pm$9.3 km~s$^{\rm {-1}}$
reported by Rutledge et al. (1997a,b). 
A value of v$^{\rm hel}_{\rm r}$ = $-$28.5$\pm$6.5 km~s$^{\rm {-1}}$
was derived by Terndrup et al. (1998),
 and v$^{\rm hel}_{\rm r}$ = $-$21.1$\pm$3.4 km~s$^{-1}$
is reported  in the compilation by Harris
(1996, updated in 2010)\footnote{http://www.physics.mcmaster.ca/Globular.html}.

Finally, the variations of radial velocities given in Table 2 are within
the errors,
indicating that there is no evidence of binarity. In particular, the mass
 transfer from an AGB companion scenario, mentioned in Sect. 1, would be
 favoured in case of such evidence.

\begin{table*}
\caption[1]{Identifications, positions, and magnitudes.
$JHK_{s}$ from both the 2MASS and VVV surveys are given. }
\small
\begin{flushleft}
\tabcolsep 0.15cm
\begin{tabular}{ccccccccccccccccccc}
\noalign{\smallskip}
\hline
\noalign{\smallskip}
\hline
\noalign{\smallskip}
{\rm star} & OGLE no.& 2MASS ID & $\alpha_{2000}$ & $\delta_{2000}$ & $V$ & $I$ & $J$ & $H$ & $K_{\rm s}$ &   $J_{\rm VVV}$ 
& {\rm H$_{\rm VVV}$} & {\rm K$_{\rm VVV}$} &  \cr
\noalign{\vskip 0.2cm}
\noalign{\hrule\vskip 0.2cm}
\noalign{\vskip 0.2cm}
B-107 &402361 &18033660-3002164 & 18:03:36.59 & -30:02:16.1 & 15.977 & 14.313 & 13.006 & 11.803 & 11.574 & 13.023 & 12.443 & 12.272 &   \cr
B-108 &245265 & ---             & 18:03:35.19 & -30:02:04.9 & 16.287 & 14.400 & ---    & ---    & ---   & ---   & --- & ---   &  \cr
B-122 &244582 &18033338-3001588 & 18:03:33.35 & -30:01:58.3 & 16.001 & 14.281 & 12.707 & 11.988 & 11.128 &  12.949    & 12.300 & 12.130   &  \cr
B-128 &402607 &18034463-3002107 & 18:03:44.62 & -30:02:10.4 & 16.260 & 14.553 & 12.860 & 12.614 & 12.438 & --- & --- & --- &  \cr
B-130 &402531 &18034102-3003036 & 18:03:41.01 & -30:03:03.0 & 16.302 & 14.642 & 12.850 & 12.307 & 10.634 & 13.359  & 12.747 & 12.525  \cr
\noalign{\smallskip} \hline \end{tabular}
\end{flushleft}
\label{starmag}
\end{table*}

\section{Photometric stellar parameters}

\subsection{Temperatures}

The $VIJHK_{\rm s}$ magnitudes are given in Table \ref{starmag}.
$V$ and $I$ data of NGC 6522 were collected
from the Optical Gravitational Lensing Experiment (OGLE) survey,
the OGLE-II release, Field Bul-SC45 from 
Udalski et al. (2002), 2MASS   $J$, $H$, and $K_{\rm s}$   
from Skrutskie et al. (2006)\footnote{
$\mathtt{http://ipac.caltech.edu/2mass/releases/allsky/}$}
and $J$, $H$, and $K_{\rm s}$ from the
 Vista Variables in the Via Lactea survey
(VVV, Saito et al. 2012).
Note that for star B-130, the NASA/IPAC IRSA data
base currently lists J=11.850,
but the previous value of J=12.850, as given in B09,
 appears to be correct.

 We adopted a reddening of $E(B-V)$ = 0.45
according to the discussion in B09.
Reddening laws from 
Dean et al.   (1978) and  Rieke  \& Lebofsky (1985), namely,  
R$_{\rm V}$ = A$_{\rm V}$/E($B-V$) = 3.1,
 E($V-I$)/E($B-V$)=1.33,
 E($V-K$)/E($B-V$)=2.744,
E($J-K$)/E($B-V$)=0.527, implying colour corrections of
A$_{\rm I}$/E($B-V$) = 1.77, A$_{\rm J}$/E($B-V$) = 0.88,
A$_{\rm H}$/E($B-V$) = 0.55, and  A$_{\rm K}$/E($B-V$) = 0.356 were adopted.

Effective temperatures were derived from  $V-I$, $V-K$, and $J-K$ using
the colour-temperature calibrations of Alonso et al.  (1999, hereafter
AAM99).
To transform $V-I$ from the Cousins to the Johnson system
 we use the relation ($V-I$)$_{C}$=0.778($V-I$)$_{J}$    (Bessell 1979).
The $J,H,K_S$ 2MASS magnitudes  and colours were  transformed from the 2MASS
system to that of the California Institute of  Technology (CIT), 
and from this to that of the
Telescopio Carlos S\'anchez (TCS),  using the relations established by
Carpenter (2001)  and  Alonso et  al.  (1998).  
The VVV $JHK_{s}$ colours were transformed to the 2MASS $JHK_{s}$ system using
relations reported by Soto et al. (2013).
 The derived photometric effective
temperatures are  listed in  Table~\ref{tabteff}. 

\subsection{Gravities}

Adopting T$_{\odot}$=5770 K, M$_{\rm bol \odot}$=4.75, M$_*$=0.85 M$_{\odot}$
and
a distance modulus of ($m-$M)$_0$ = 13.91, A$_V$ = 1.72 (Barbuy et al. 1998),
 the following  classical  relation was used to derive gravities:

\[
\log g_*=4.44+4\log \frac{T_*}{T_{\odot}}+0.4(M_{\rm bol*}-M_{\rm bol{\circ}})+\log \frac{M_*}{M_{\odot}} 
\].

 The   bolometric  corrections from AAM99 and
corresponding gravities are given in Table~\ref{tabteff}.


\section{Spectroscopic stellar parameters}

The  equivalent  widths (EW) were measured  using  the  automatic  code
DAOSPEC that was developed by Stetson \& Pancino
 (2008). 

The EWs measured for the  \ion{Fe}{I} and \ion{Fe}{II} 
 lines are reported in Table A.5, where they are compared 
with values from B09. We restrained the list of Fe lines to those
in the region 6100-6800 {\rm \AA} for the usual reasons of a higher
S/N ratio and a lower line crowding in this wavelength region, 
than in the region 4800-6100 {\rm \AA}. Another reason
was the possibility of a more straightforward comparison of differences
in the stellar parameters of B09.
 For star B-108
we also measured the lines of star B-108 using IRAF for a comparison.
 In Fig. \ref{residuals} we compare EWs from B09 and the present
measurements, as well as their residuals. 
The EWs for stars B-107, B-122, B-128,
and B-130 agree well within $\pm$20m{\rm \AA}. For B-108 the values
show a large scatter, which can be explained 
recalling that for this object two stars were
included in the slit, as explained in Sect. 5.2.
We also searched for a possible trend between EW and wavelength,
but found none, except for lines in the region 6700-6800 {\rm \AA},
but since we only used lines with  30 $<$ EW $<$ 100 m{\rm \AA},
only very few EWs for lines in this region were used.

 The error in EWs as given by Cayrel (1988) and Cayrel et al. (2004) is 
$\sigma={1.5 \over S/N} \sqrt{FWHM*\delta{x}}$.
We measured a mean FWHM = 12.5 pixels, or 0.184 {\rm \AA}.
 The CCD pixel size is 15 $\mu$m, or $\delta{x}$ = 0.0147 {\rm \AA}
in the spectra. By assuming a mean
S/N=100, we derive an error $\Delta$EW $\sim$ 0.8 m{\rm \AA}
(note that this formula neglects the uncertainty in 
the continuum placement).

For the \ion{Fe}{I} line list given in Table A.5,
we employed the respective oscillator strengths
 described in Zoccali et al. (2004) and Barbuy et al. (2006, 2007).

Photospheric 1D models for the sample giants  were extracted from the
MARCS model atmosphere grid (Gustafsson   et al. 2008).
The LTE  abundance  analysis and  the spectrum  synthesis calculations
were performed using the code described  in Spite (1967), Cayrel  et al.
(1991), Barbuy et al. (2003), and Coelho et al. (2005). An iron abundance
of $\epsilon$(Fe)=7.50 (Grevesse \& Sauval 1998) was adopted.
Molecular lines
of CN  (A$^2$$\Pi$-X$^2$$\Sigma$), C$_2$ Swan (A$^3$$\Pi$-X$^3$$\Pi$), TiO
(A$^3$$\Phi$-X$^3$$\Delta$) $\gamma$,   and  TiO (B$^3$$\Pi$-X$^3$$\Delta$)
$\gamma$' systems  were taken   into  account.

The stellar   parameters  were  derived    by initially  adopting  the
photometric effective  temperature  and  gravity,  and   then  further
constraining  the  temperature by imposing an excitation equilibrium for
the \ion{Fe}{I} lines.
Five  \ion{Fe}{II} lines were measurable, and their respective
oscillator strengths from Bi\'emont et al. (1991) were renormalized by
Mel\'endez \&   Barbuy (2009). From this we derived gravities
by imposing ionization equilibrium. 
Microturbulence velocities v$_t$  
were  determined by canceling the trend of \ion{Fe}{I} abundance vs.  
equivalent width.

\begin{figure}
\centering
\psfig{file=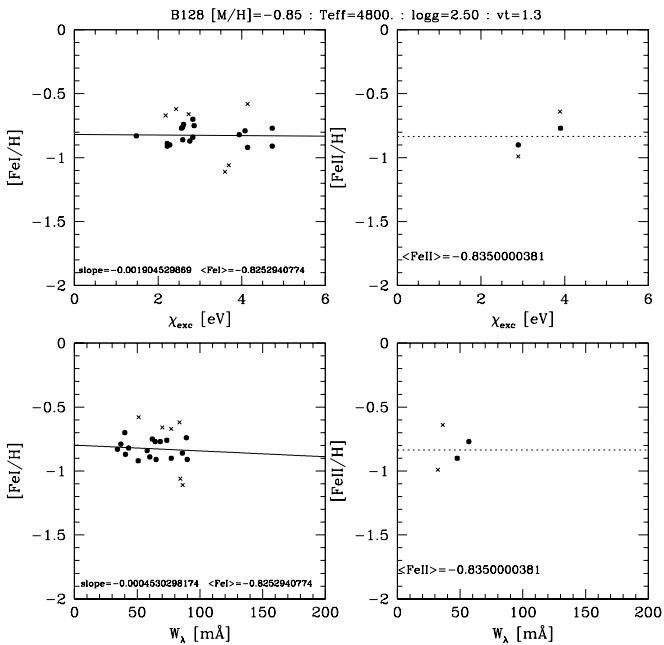,angle=0.,width=9.0 cm}
\caption{Excitation and ionization equilibria of \ion{Fe}{I} and \ion{Fe}{II} lines for star B-128.
}
\label{abon} 
\end{figure}

The final spectroscopic parameters T$_{\rm eff}$, log g, [\ion{Fe}{I}/H],
 [\ion{Fe}{II}/H],  [Fe/H], and the
 v$_t$ values  are reported in  the  last columns
  of Table~\ref{tabteff}. An example of
excitation and ionization equilibria using
the \ion{Fe}{I} and \ion{Fe}{II} lines
is shown in Fig. \ref{abon} for star B-128.
 We preferred to use the spectroscopic parameters for consistency 
to fulfill the  excitation and ionization equilibria of the Fe lines,
 even if overionization of
Fe is expected, which might change these equilibria somewhat. The difference
in  abundances between \ion{Ti}{I} and  \ion{Ti}{II} of 0.13 dex 
can be interpreted
as having a too high effective temperature.
 We found spectroscopic temperatures  systematically hotter
by 200 K  than the photometric temperatures
(e.g. Zoccali et al. 2004).

 In Table \ref{tabteff} a second line
is added for each star, that lists the stellar parameters
derived from GIRAFFE spectra in B09. The agreement is excellent,
considering that the two derivations were made entirely independent
of each other,
with the only constraint of using the same list of Fe lines 
(with EWs measured from GIRAFFE and UVES).

\begin{table*}
\caption[1]{
Photometric stellar parameters derived using the calibrations by Alonso et al. (1999)
 for $V-I$, $V-K$, $J-K$, bolometric corrections, and bolometric magnitudes
and the corresponding gravities log $g$,
and final spectroscopic parameters.
For each star the last columns in the first line give
the spectroscopic parameters from UVES spectra (present work), and
the second line reports those
derived from GIRAFFE spectra in B09.} 
\small
\begin{flushleft}
\begin{tabular}{c@{}c@{}c@{}c@{}c@{}c@{}c@{}c@{}cc@{}c@{}c@{}c@{}c@{}c@{}c}
\noalign{\smallskip}
\hline
\noalign{\smallskip}
\hline
\noalign{\smallskip}
& \multicolumn{4}{c}{\hbox{}} Photometric\phantom{-} parameters & & & & \multicolumn{7}{c}{\hbox{}} Spectroscopic\phantom{-} parameters\\
\cline{2-9}  \cline{11-16}    \\
{\rm star} & T($V-I$) & \phantom{-}\phantom{-}T($V-K$) & \phantom{-}\phantom{-}T($J-K$)
 &\phantom{-}\phantom{-}T($V-K$)  &  \phantom{-}\phantom{-}T($J-K$) &  
\phantom{-}\phantom{-}${\rm BC_{V}}$ & \phantom{-}\phantom{-}${\rm M_{bol}}$ &
\phantom{-}\phantom{-}log g & & \phantom{-}T$_{\rm eff}$ &
\phantom{-}\phantom{-}log g &\phantom{-}\phantom{-}[FeI/H] & 
\phantom{-}\phantom{-}[FeII/H] &
 \phantom{-}\phantom{-}[Fe/H] 
& ${\rm v_t }$ \\
 &  & 2MASS & 2MASS  & VVV  &VVV  & &  & & &  &  &    \\
 & (K) & (K) &  (K) &  (K) & (K) & & & & & (K) & & & & & km~s$^{-1}$  \\
\noalign{\smallskip}
\noalign{\hrule}
\noalign{\smallskip}
{B-107} & 4594 & --- & --- & 4664 & 4860 & $-0.37$ & \phantom{-}$0.97$ & \phantom{-}2.50 & &  4990 & \phantom{-}2.00 & $-$1.11  & $-$1.14 & $-$1.12 & 1.40 \\
 & & & & & & & & & &  4900 & \phantom{-}2.10 & $-$1.15  & $-$1.06 & $-$1.11 & 1.40 \\
{B-122} & 4487 & ---  & --- & 4527 & 4621  & $-0.41$ & \phantom{-}$1.04$ & \phantom{-}2.48 & &  4900 & \phantom{-}2.7 & $-$0.80 & $-$0.82 & $-0.81$ & 1.55 \\
 & & & & & & & & & & 4800 & \phantom{-}2.6 & $-$0.87 & $-$0.87 & $-0.87$ & 1.10 \\
{B-128} & 4511 & 4572 & 6838 & ---&---& $-0.40$ & \phantom{-}$1.28$ & \phantom{-}2.58 & &  4800 & \phantom{-}2.5 & $-$0.81  & $-$0.82 & $-0.81$ &  1.25 \\
 & & & & & & & & & & 4800 &\phantom{-}2.7 & $-$0.79 & $-$0.79 & $-0.79$ &1.30 \\
{B-130} & 4602 & 4612  & 4773 & 4605 & 4586  & $-0.36$ & \phantom{-}$1.28$ & \phantom{-}2.62 & & 4850 & \phantom{-}2.2 & $-$1.03  & $-$1.05 & $-1.04$ & 1.45 \\
& & & & & & & & & & 4800 &\phantom{-}2.3 & $-$1.09& $-$1.10 & $-1.09$ & 1.40 \\
\noalign{\smallskip} \hline \end{tabular}
\end{flushleft}
\label{tabteff}
\end{table*}

\subsection{Solar and Arcturus abundances}

The lines were checked in the solar spectrum observed with the same 
instrumentation as the sample 
stars\footnote{http://www.eso.org/\-observing/\-dfo/\-quality/\-UVES/\-pipeline/solar{$_{-}$}spectrum.html}
and in the Arcturus spectrum (Hinkle et al. 2000). 
The solar abundances from the literature are given in Table \ref{sol}. We
adopted the  Grevesse et al. (1998) abundances, which
are similar to the latest abundances by Grevesse et al. (2014)
 or those by Lodders (2009). 
Literature parameters and abundances in Arcturus
($\alpha$ Boo, HR 5340, HD 124897, HIP 69673) are given in Table 6, 
with the parameters and abundances of the elements
derived here for Arcturus reported in the last line of 
Table 6. 


\begin{table}
\caption{Solar {\it r}- and {\it s}-process fractions from 
Simmerer et al. (2004), and solar abundances 
from 1: Kur\'ucz (1993); 
2: Grevesse et al. 1998; 3: Asplund et al. 2009; 4: Lodders 2009;
 5: Grevesse et al. 2014. Note: A(X) = log(X/H)+12.}
\label{sol}      
\centering                          
\begin{tabular}{l@{} l l l l l l l l} 
\hline\hline                 
\hbox{El.} & \hbox{Z} & \multicolumn{2}{c}{Fraction} 
& \multicolumn{5}{c}{\hbox{A(X)$_{\odot}$}}  \\
\cline{3-4} \cline{5-9}   \\
 &  & r & s & (1) & (2)  & (3) & (4) & (5) \\
\hline                        
\hbox{C} &   6 &  --   &  --   & 8.52 &  8.55  &  8.43  &  8.39  & --- \\
\hbox{N} &   7 &  --   &  --   & 8.01 &  7.97  &  7.83  &  7.86  & --- \\
\hbox{O} &   8 &  --   &  --   & 8.89 &  8.87  &  8.69  &  8.73  & --- \\
\hbox{Na} & 11 &  --   &  --   & 6.29 &  6.33  &  6.24  &  6.30  & --- \\
\hbox{Mg} & 12 &  --   &  --   & 7.54 &  7.58  &  7.60  &  7.54  & --- \\
\hbox{Al} & 13 &  --   &  --   & 6.43 &  6.47  &  6.45  &  6.47  & --- \\
\hbox{Si} & 14 &  --   &  --   & 7.51 &  7.55  &  7.51  &  7.52  & --- \\
\hbox{Ca} & 20 &  --   &  --   & 6.32 &  6.36  &  6.34  &  6.33  & --- \\
\hbox{Ti} & 22 &  --   &  --   & 4.95 &  5.02  &  4.95  &  4.90  & --- \\
\hbox{Fe} & 26 &  --   &  --   & 7.63 &  7.50  &  7.50  &  7.45  & 7.45 \\
\hbox{Sr} & 38 & 0.110 & 0.890 & 2.86 &  2.97  &  2.87  &  2.92  & 2.83 \\
\hbox{Y}  & 39 & 0.281 & 0.719 & 2.20 &  2.24  &  2.21  &  2.21  & 2.21 \\
\hbox{Zr} & 40 & 0.191 & 0.809 & 2.46 & 2.60  &  2.58  &  2.58  & 2.59 \\
\hbox{Ba} & 56 & 0.147 & 0.853 & 2.09 &  2.13  &  2.18  &  2.18  & 2.25 \\
\hbox{La} & 57 & 0.246 & 0.754 & 1.18 &  1.17  &  1.10  &  1.14  & --- \\
\hbox{Eu} & 63 & 0.973 & 0.027 & 0.47 &  0.51  &  0.52  &  0.52  & 0.52 \\
\hline                                  
\end{tabular}
\end{table}
 
\begin{table*}
\small
\label{arcturus}
\caption[4]{
   Parameters employed for Arcturus, which was used as a reference star.
References: 1 - M\"ackle et al. (1975); 2 - McWilliam \& Rich (1994); 
3 - Peterson et al. (1993);
4 - Gopka et al. (2001); 
5 - Mishenina et al. 2001, 2003;  
6 - Luck \& Heiter 2005; 
7 - parameters from Mel\'endez et al. 2003 and abundances from the present work.
 }
\begin{flushleft}
\begin{tabular}{lrrrrrrrrrrrrrrrrrrrrrrrrrrrrrrrrrrr}
\noalign{\smallskip}
\hline
\noalign{\smallskip}
T$_{\rm eff}$&\phantom{-} log g  & \phantom{-}v$_{\rm t}$ & [Fe/H] 
 & [O/Fe] & [Ca/Fe]  & [Sr/Fe]  & [Y/Fe]  & [Zr/Fe]  & [Ba/Fe] & [La/Fe]
 & [Eu/Fe] & ref. & \\
\noalign{\vskip 0.2cm}
\noalign{\hrule\vskip 0.2cm}
\noalign{\vskip 0.2cm}
4260 & \phantom{-}0.90 &--- &\phantom{-}-0.70 &+0.10 &+0.25 &-0.40 &-0.30 &-0.20 &-0.25 &-0.20 &-0.15 & 1 & \\
4280 & \phantom{-}1.30 &1.40 &\phantom{-}-0.54 &--- &+0.17
 &--- & -0.42 &-1.07 &+0.16 &-0.15
 &+0.32 & 2 & \\
4300 & \phantom{-}1.50 &  1.70 &\phantom{-}-0.50 &+0.40 &+0.30 & 0.0 &0.0 &0.0&0.0 &0.0
 &0.0 & 3 & \\
4350 & \phantom{-}1.60 &  1.60 &\phantom{-}-0.56 & --- &---
 &-0.27 &-0.08 &-0.07 &--- &-0.15 &+0.15 & 4 & \\
4350 & \phantom{-}1.60 & 1.60 &\phantom{-}-0.58 &--- &+0.26 &-0.17 &-0.01 &--- &-0.07 &-0.02 &+0.39 & 5 & \\
4340 & \phantom{-}1.93 &  1.87 &\phantom{-}-0.55 &+0.60
 &-0.01 &--- &-0.07 &--- &-0.28 &-0.01 &+0.21 & 6 & \\
4275 & \phantom{-}1.55 &  1.65 &\phantom{-}-0.54 &+0.39 &+0.28 & -0.3 &-0.3 &-0.07 &-0.30 &-0.30 &+0.15 & 7 & \\
\noalign{\smallskip} \hline \end{tabular}
\end{flushleft}
\end{table*}

\subsection{Star B-108}\label{b108}

The star B-108 did not easily converge
in terms of spectroscopic parameters. Therefore
we investigated the star B-108 in detail, which 
was reported to have large heavy-element
enhancements in C11.

\begin{figure}
\centering
\psfig{file=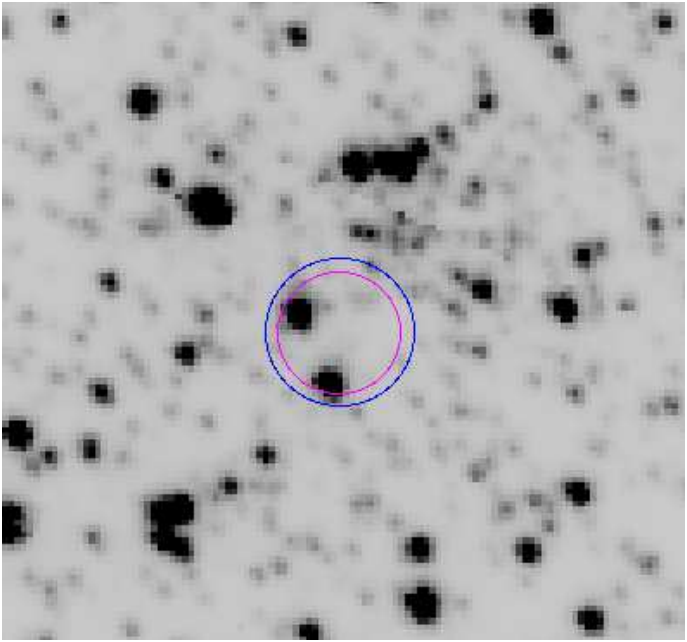,angle=0.,width=9.0 cm}
\caption{ACS image of B-108. The circles indicate the size
of the UVES and GIRAFFE fibers: two stars are present. 
}
\label{ACS} 
\end{figure}

We show in Fig. \ref{ACS} the Advanced Camera for Surveys (ACS)
of the Hubble Space Telescope image,  taken on 10/07/2003, 
and superposed the size 
 of the fibers of UVES and GIRAFFE,
which are $1\farcs0$ and   $1\farcs2$. 

\begin{figure}
\centering
\psfig{file=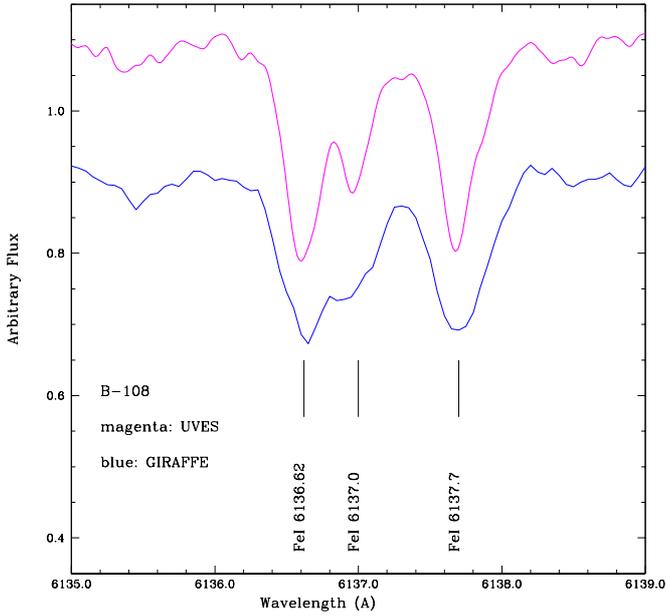,angle=0.,width=9.0 cm}
\caption{UVES and GIRAFFE spectra of star B-108.
The lines show two components of superposed spectra
of the two stars. The two components are indicated by the
asymmetry of the line profiles. 
}
\label{b108comp} 
\end{figure}

The sample star B-108 has no reported proper motions in the OGLE-II
or in the UCAC4 (Zacharias et al. 2013) data bases.
For the other sample stars the proper motions are
 of the order of 0.005 arcsec/yr, which are
low enough not to change its location relative to the fiber size.
From the ACS image it appears that the OGLE-II coordinates of star B-108 
are shifted relative to its location. This is a warning not
 to use the coordinates directly but to verify images when possible.
The image also shows that there is a blend of two stars. 
We examined the spectra in more detail, 
and in fact the asymmetry of lines reveals
 spectra of two stars of similar radial velocity, but with a small difference,
as exemplified in Fig. \ref{b108comp}.

The lower resolution of GIRAFFE caused a misanalysis of star B-108 
in B09, while the UVES
higher resolution which permits a better deblending of the spectra of the
two stars, led to a less satisfactory set of parameters.

 B-108 was finally discarded from the sample because o
a clear contamination of its spectrum, and previous abundances
of this star cannot be considered. It would be
 useful to observe both
 of these blended stars and determine whether the abundance
 enhancements detected in C11 are present.

\section{Abundance ratios}

Abundances ratios  were   obtained  by means of line-by-line  spectrum
synthesis calculations that were compared with the observed spectra,
for the line lists given in Tables~ 8, 9, and  11. 

\subsection{Carbon, nitrogen, and oxygen}

\begin{figure}
\centering
\psfig{file=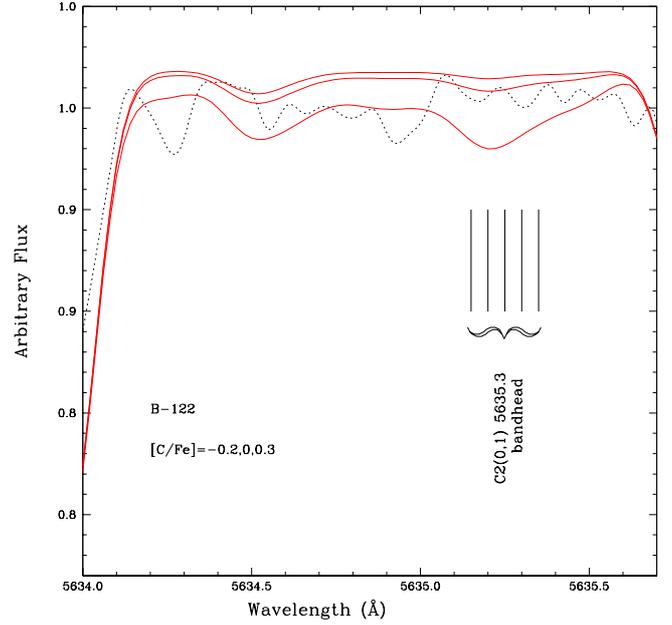,angle=0.,width=9.0cm}
\caption{C$_{2}$(1,0) bandhead at 5635.3 {\rm \AA} in B-122. }
\label{b122c2} 
\end{figure}
\makeatletter
\renewcommand{\table}[1][]{\@float{table}[!htp]}
\makeatother 
\begin{figure}
\centering
\caption{Star B-128: CN(5,1) bandhead at 6332.18 {\rm \AA}. }
\label{cn51} 
\psfig{file=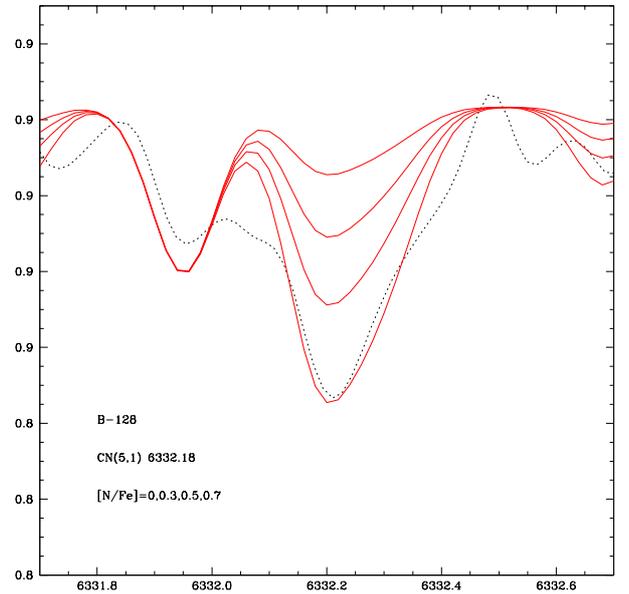,angle=0.,width=9.0cm}
\end{figure}

\begin{figure}
\centering
\caption{[Na/Fe] vs. [O/Fe] for the sample stars compared
with stars of NGC 6121. Symbols: blue filled triangles: stars of NGC 6121;
(Na, O) for the sample stars: a) green filled squares: [Na/Fe] not corrected,
b) open red squares: [Na/Fe] corrected.  }
\label{na-o} 
\psfig{file=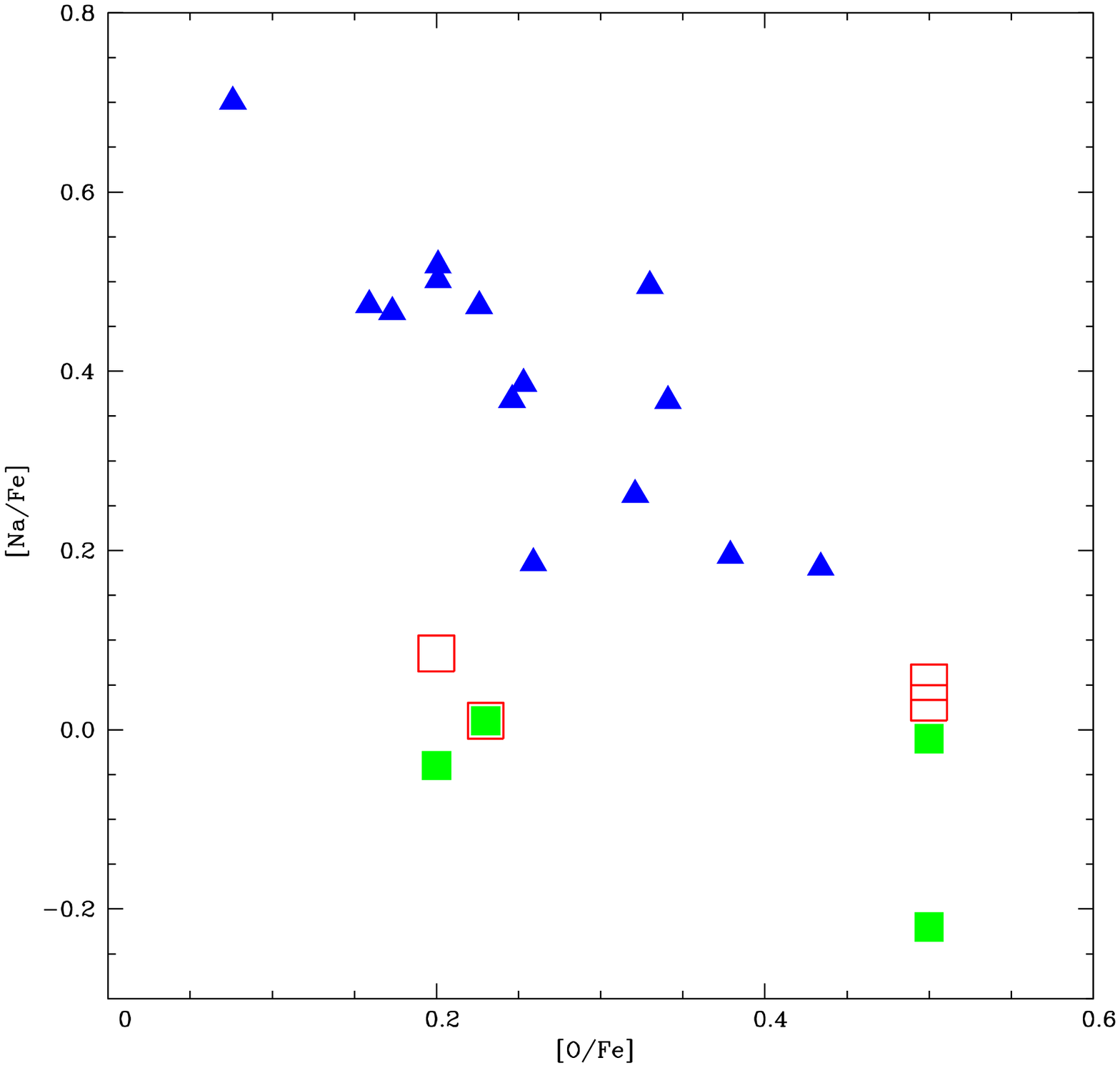,angle=0.,width=9.0cm}
\end{figure}

The carbon abundances were estimated from the C$_{2}$(0,1) bandhead
at 5635.3 {\rm \AA}. The fit to B-122 is shown in Fig. \ref{b122c2}.
 Since the bandhead is extended, as can be seen
in Fig. \ref{b122c2}, a mean
value was deduced from the overall fit, but giving more weight
to the bandhead.
 The list of C$_2$ lines is the laboratory list from Phillips \&
Davis (1968) The resulting C and N abundances are given in Table
\ref{c2cn}. 

 The forbidden oxygen line [OI] 6300.311 {\rm \AA} 
 in B-128 and B-130 can be detected on the right side of a telluric line.
The same is true for B-122, but the oxygen line is weaker.
To examine this we overplotted the spectrum of a B star with high
rotation that shows a continuum with the telluric lines in it.
 For B-107 there is no blend with a telluric line, but the oxygen line
appears anomalously strong.

The  [OI] 6363.776 {\rm \AA} line has the advantage of not being masked
by telluric lines, but the disadvantage of being weaker and blended
on the blue side by a FeI line.

The nitrogen abundances were measured using the CN (5,1) 6332.18 {\rm \AA}
and CN (6,2) 6478.48 {\rm \AA} of the CN A$^{2}\Pi$-X$^{2}\Sigma$ red system.
The CN(5,1) bandhead in star B-128 is shown in Fig. \ref{cn51}.
 
We iteratively recomputed the abundances in a sequence C, O, and N, 
because
the N abundance is very sensitive tothe C abundance in CN molecules, and O
is sensitive to C abundances, because of the CO formation.

The results show a low carbon [C/Fe]$<$0.1 for all stars, 
enhanced nitrogen, as expected
in giants, and enhanced oxygen 0.2$<$[O/Fe]$<$0.5, 
typical of enrichment by SNe type II.

\begin{table}
\begin{flushleft}
\caption{C and N abundances from C$_2$ and CN bandheads.}
\label{c2cn}      
\centering 
\small         
\begin{tabular}{lccccccccccc}     
\noalign{\smallskip}
\hline\hline    
\noalign{\smallskip}
\noalign{\vskip 0.1cm} 
species & {\rm $\lambda$} ({\rm \AA})
& \hbox{B-107} &\hbox{B-122} &\hbox{B-128} &\hbox{B-130}   \\
\noalign{\vskip 0.1cm}
\noalign{\hrule\vskip 0.1cm}
\noalign{\vskip 0.1cm}
C$_{2}$(0,1) & 5635.3 & $\sim$0.0 & $\leq$$-$0.2 &  $\sim$0.1 & $\sim$0.0 \\ 
CN(5,1) & 6332.16 & --- & $\leq$0.7 &  $\sim$0.7 & $\sim$0.7: \\ 
CN(6,2) & 6478.48 & --- & $\leq$0.7 &  $\sim$0.5 & $\sim$0.7 \\ 
\noalign{\smallskip} \hline \end{tabular}
\end{flushleft}
\end{table}

\subsection{Odd-Z elements Na, Al, and alpha elements}

In Table \ref{tablines} we report the line-by-line
$\alpha$-element and odd-Z element abundances for the same line list
as in B09, which essentially agrees well for
 the two abundance derivations. In Table \ref{newlist}
we report a complementary list of lines in the wavelength region
4800 - 6100 {\rm \AA}, now available with the UVES spectra.
This bluer line region is subject to more blending lines however,
and is therefore less reliable.
We list oscillator strengths from the literature and the adopted values
based on fits to the solar spectrum observed with the UVES spectra
as used in Sect. 3.1.

Table \ref{alpha} lists the final abundance results from Tables
 \ref{tablines} and \ref{newlist}.

{\it Odd-Z elements Na, Al}

 We inspected the abundance results as a function
of effective temperature and microturbulence velocity.
 The abundances of odd-Z elements Na and Al show a trend vs.
effective temperature, which was corrected for. The corrected
values are given in parenthesis in Table \ref{alpha}.
For the other elements there is no evidence of a significant trend.

{\it Alpha elements}

The 'bona-fide' $\alpha$-elements in terms of nucleosynthesis
production by SNe type II, O, and Mg, are enhanced in the
four sample stars. Silicon is moderately enhanced, and Ca is moderately
enhanced in two stars and is solar in one star. As for Ti, TiII gives
moderate enhancements, while TiI gives solar ratios, and
for both we find a mean low enhancement of +0.05 $<$ [Ti/Fe] $<$ +0.1.

The results in B09 of
[O/Fe] = +0.4, [Mg/Fe] = +0.3, [Si/Fe] = +0.3, [Ca/Fe] = +0.2, [Ti/Fe] = +0.3,
are the same as ours for O and Mg, but we find lower enhancements
for Si, Ca, Ti, of [Si/Fe]=+0.15, [Ca/Fe]=+0.10 and [Ti/Fe]=+0.15.

{\it Na-O anticorrelation}

 A Na-O anticorrelation was clearly made evident in 
many globular clusters, most likely caused by material processed
in the Ne-Na cycle of proton capture reactions. The expectation
is that O is depleted and Na and Al are enhanced.
(e.g. Gratton et al. 2012
and references therein). For this reason,
 we compared the present Na and O abundances with
those for the globular cluster NGC 6121,
with a  metallicity of [Fe/H]=-1.16, which is similar to NGC 6522. 
The values of [Na/Fe] vs. [O/Fe] with abundances
derived for 14 stars of NGC 6121 by Carretta et al. (2009) are plotted in 
Fig. \ref{na-o}. The present results are plotted with the original
Na abundance derivation, and the values were corrected as explained above.

The four sample stars have [Na/Fe]$\sim$0 and
 do not seem to show a Na-O anticorrelation,
differently from NGC 6121 and many other clusters, as shown
by Carretta et al. (2009). More stars are needed 
to arrive at definitive conclusions.
 Since the stars in our sample share the same Na enrichment,
 we may conclude that they belong to the same generation within
 the cluster. The rather low  [Na/Fe] ratio 
 indicates that probably they all belong to a first generation
of stars.

\begin{table*}
\caption{Abundance ratios [X/Fe] of alpha-elements O, Mg, Si, Ca, Ti,
 and odd-Z Na, Al, and atomic parameters adopted.
Column 4 reports the log gf adopted  as described in B09.
 For each star the left column reports the results from B09
 and the right column gives the present results.
}
\label{tablines}
\begin{flushleft}
\begin{tabular}{llllrrrrrrrrrrrrrrrrr}
\noalign{\smallskip}
\hline\hline    
\noalign{\smallskip}
\noalign{\vskip 0.1cm} 
Species & \hbox{\rm $\lambda$}     & \hbox{\rm $\chi_{ex}$} & \hbox{log~gf} & & & & & & [X/Fe] & & & &    \\
   & ({\rm \AA}) & (eV)  &  & {B-107}  &   {B-107b}
 & {B-122} & {B-122b} &  {B-128} & {B-128b} & {B-130} &  {B-130b} \\
\noalign{\smallskip}
\hline\hline    
\noalign{\smallskip}
\noalign{\vskip 0.1cm} 
OI & 6300.311 & 0.00   & $-$9.716       & ---  &  $>$0.5: & ---   &  +0.3:  & ---  & +0.30   & +0.5: & +0.50 \\ 
OI & 6363.776 & 0.00    & $-$10.25      &  +0.5: & +0.5: & +0.7: & +0.2   & +0.7: & +0.15  & ---   & +0.50      \\
NaI & 6154.230 & 2.10  & $-$1.56        &   $-$0.30  &$-$0.20  & +0.10  
&$-$0.10  & +0.10  & \phantom{+}0.00 & +0.10   & \phantom{+}0.00 \\
NaI & 6160.753 & 2.10   &  $-$1.26      &   $-$0.30  &\phantom{+}0.00 &
 +0.20  &$-$0.10  & +0.10  &+0.10  & +0.20   &+0.05 \\
MgI & 6318.720 & 5.11  &$-$2.10        &   +0.20  &+0.30 & +0.20   &+0.20 & +0.20   &+0.30 & +0.40    & +0.30 \\
MgI & 6319.242 & 5.11   &$-$2.36       &   ---    &+0.30 & +0.20   &+0.10 & +0.20   &+0.30 & +0.40    & +0.30 \\
MgI & 6319.490 & 5.11    &$-$2.80      &   +0.20  &+0.30 & +0.20   &+0.10 & +0.40   &+0.40 & +0.40    & ---\\
MgI & 6765.450 & 5.75  &  $-$1.94       &   +0.40 &--- & ----   &--- & +0.20  &0.00 & +0.40   & --- \\
SiI & 6142.494 & 5.62   & $-$1.50       &   +0.20   &+0.20 &  
\phantom{+}0.00   & \phantom{+}0.00 & +0.20   &+0.30 & +0.40    & ---\\
SiI & 6145.020 & 5.61  & $-$1.45        &   +0.20   &+0.05 & +0.20   &
\phantom{+}0.00 & +0.20   &+0.30 &  ---    & ---\\
SiI & 6155.142 & 5.62  & $-$0.85        &   +0.1   &+0.05 & $-$0.1   &
\phantom{+}0.00 & +0.20   &+0.15 & +0.30    &+0.15\\
SiI & 6237.328 & 5.61   &  $-$1.01      &   +0.20   &+0.20 & +0.20   &+0.20 & +0.30   &+0.10 & +0.40    & +0.15\\
SiI & 6243.823 & 5.61   & $-$1.30       &   +0.20   &+0.20 & +0.30   &
\phantom{+}0.00 & +0.4   &\phantom{+}0.00 & +0.40    & +0.15\\
SiI & 6414.987 & 5.87    & $-$1.13      &   +0.20   &+0.20 & +0.20   &0.00 & +0.20   &+0.20 & +0.20    & +0.15\\
SiI & 6721.844 & 5.86  & $-$1.17        &   +0.30   &\phantom{+}0.00 &  ---   
&\phantom{+}0.00 & +0.20  &+0.30 & +0.40   & +0.15\\
CaI & 6156.030 & 2.52   &$-$2.39       &   +0.30   &+0.30 &  \phantom{+}0.00   &+0.30 & +0.30   &+0.30 & ---     & +0.30\\
CaI & 6161.295 & 2.51   & $-$1.02       &    \phantom{+}0.00   &+0.15 & +0.30   
&\phantom{+}0.00 & +0.20   &+0.30 & +0.30    & +0.20 \\
CaI & 6162.167 & 1.89   & $-$0.09       &   +0.20   &+0.15 &  \phantom{+}0.00   &
\phantom{+}0.00 &  \phantom{+}0.00   &\phantom{+}0.00 & +0.10    & \phantom{+}0.00\\
CaI & 6166.440 & 2.52  & $-$0.90        &    \phantom{+}0.00   &
\phantom{+}0.00 & +0.30   &$-$0.05 &  \phantom{+}0.00   &\phantom{+}0.00 & +0.30    & \phantom{+}0.00\\
CaI & 6169.044 & 2.52  & $-$0.54        &   +0.10   &+0.25 & +0.30   &$-$0.05 & +0.30   &+0.30 & +0.10    & \phantom{+}0.00\\
CaI & 6169.564 & 2.52   & $-$0.27       &   +0.20   &+0.10 & +0.30   &\phantom{+}0.00 & +0.30   &+0.30 & +0.20    & \phantom{+}0.00\\
CaI & 6439.080 & 2.52   & +0.3        &   +0.30   &+0.25 & +0.30   &\phantom{+}0.00 & +0.30   &+0.30 & +0.30    & +0.30\\
CaI & 6455.605 & 2.52  & $-$1.35        &    \phantom{+}0.00   &+0.10 & +0.30   &+0.10 & $-$0.10   &+0.10 & +0.40   &+0.30\\
CaI & 6464.679 & 2.52   &$-$2.10       &    \phantom{+}0.00   &+0.30 & +0.30   &+0.30 & +0.30  &+0.30 &  ---    & ---\\
CaI & 6471.668 & 2.52  & $-$0.59        &    \phantom{+}0.00   &
\phantom{+}0.00 & +0.30   &\phantom{+}0.00 & +0.30   &+0.30 & +0.30    & +0.20\\
CaI & 6493.788 & 2.52     & \phantom{+}0.00       &   $-$0.30   &+0.15 &  \phantom{+}0.00   &\phantom{+}0.00 &  \phantom{+}0.00   &+0.30 & +0.10    &+0.30\\
CaI & 6499.654 & 2.52     & $-$0.85     &   $-$0.30   &--- &  \phantom{+}0.00   &--- &  \phantom{+}0.00   &--- & +0.10    & ---\\
CaI & 6572.779 & 0.00     & $-$4.32     &    \phantom{+}0.00   &+0.10 & +0.20   &\phantom{+}0.00 &  \phantom{+}0.0   &\phantom{+}0.00 & +0.20    & +0.10\\
CaI & 6717.687 & 2.71      & $-$0.61    &    \phantom{+}0.00   &+0.15 & +0.30   &
\phantom{+}0.00 & +0.30   &+0.20 & +0.30    & +0.10\\
TiI & 6126.224 & 1.07     & $-$1.43     &   +0.20   &+0.10 & +0.10   &\phantom{+}0.00 & +0.10   &\phantom{+}0.00 & +0.20    & \phantom{+}0.00\\
TiI & 6258.110 & 1.44     & $-$0.36     &    \phantom{+}0.00   &+0.15 & +0.20   &
\phantom{+}0.00 & +0.20   &\phantom{+}0.00 &  \phantom{+}0.00    & \phantom{+}0.00\\
TiI & 6261.106 & 1.43      & $-$0.48    &    \phantom{+}0.00   &0.00 & +0.30   &\phantom{+}0.00 &  \phantom{+}0.00   &0.00 &  \phantom{+}0.00    & \phantom{+}0.00\\
TiI & 6303.767 & 1.44     & $-$1.57     &    \phantom{+}0.00   &\phantom{+}0.00 & +0.20  &+0.30 & +0.20  &+0.30 &  ---    &  \phantom{+}0.00\\
TiI & 6336.113 & 1.44     & $-$1.74     &    ---   &\phantom{+}0.00 & +0.30   &\phantom{+}0.00 &  \phantom{+}0.00  &--- &  ---    & ---\\
TiI & 6554.238 & 1.44     & $-$1.22     &   +0.20   &--- & +0.20   &\phantom{+}0.00 & +0.20   
&\phantom{+}0.00 &  \phantom{+}0.00    & \phantom{+}0.00\\
TiI & 6556.077 & 1.46     & $-$1.07     &   +0.20   &--- &  \phantom{+}0.00   &
\phantom{+}0.00 & +0.20   &\phantom{+}0.00 & +0.40    & \phantom{+}0.00\\
TiI & 6599.113 & 0.90     &$-$2.09     &   +0.30  &--- & +0.20   &+0.20 & +0.20 &+0.20 & +0.40   & +0.20\\
TiI & 6743.127 & 0.90     & $-$1.73     &   +0.10   &\phantom{+}0.00 &  \phantom{+}0.00   &
\phantom{+}0.00 &  \phantom{+}0.0   &\phantom{+}0.00 & +0.20    & \phantom{+}0.00\\
TiII& 6491.580 & 2.06     &$-$2.10     &   +0.30   &+0.40 & +0.30   &+0.30 & +0.30   &+0.30 & +0.30    & +0.30 \\
TiII& 6559.576 & 2.05     &$-$2.35     &    \phantom{+}0.00   &+0.30 & +0.30   &+0.30 & +0.30   &+0.30 & +0.30    & +0.30\\
TiII& 6606.970 & 2.06     &$-$2.85     &   +0.20  &+0.20 & ---    &\phantom{+}0.00  & +0.30 &+0.30  & +0.30   & ---\\
\noalign{\smallskip} 
\hline 
\end{tabular}
\end{flushleft}
\end{table*}

\begin{table*}
\begin{flushleft}
\caption{Central wavelengths from NIST or Kur\'ucz line lists and 
total oscillator strengths
from line lists by Kur\'ucz, NIST, and VALD, the literature, and adopted
values. In Col. 7 literature oscillator strength values are from the
references: 1 Spite et al. (1987) and 2 Wood et al. (2013).
 Line-by-line abundances for the four sample stars are given.}
\label{newlist}      
\centering 
\small         
\begin{tabular}{l@{}ccc@{}c@{}c@{}c@{}c@{}c@{}ccccccccc}     
\noalign{\smallskip}
\hline\hline    
\noalign{\smallskip}
\noalign{\vskip 0.1cm} 
species & {\rm $\lambda$} ({\rm \AA}) & \phantom{-} \phantom{-}{\rm $\chi_{ex}$ (eV)} & 
\phantom{-}{\rm gf$_{Kurucz}$} &
 \phantom{-}{\rm gf$_{NIST}$} & \phantom{-}{\rm gf$_{VALD}$} &
 \phantom{-}{\rm gf$_{literature}$} & \phantom{-}{\rm gf$_{adopted}$}&
& \phantom{-}\phantom{-}\hbox{B-107} &\hbox{B-122} &\hbox{B-128} &\hbox{B-130}   \\
\noalign{\vskip 0.1cm}
\noalign{\hrule\vskip 0.1cm}
\noalign{\vskip 0.1cm}
NaI     & 5682.6333 & 2.102439 & $-$0.700 &$-$0.706  & $-$0.860 &$-$0.66$^1$ & $-$0.706 &
& $-$0.30  & \phantom{+}0.00 & $-$0.05 & $-$0.10  \\
NaI     & 5688.2046 & 2.104571 & $-$0.450 &$-$0.452  & $-$0.320 &$-$0.28$^1$ & $-$0.45 &
& $-$0.30  & \phantom{+}0.00   & \phantom{+}0.00 & \phantom{+}0.00 \\
NaI     & 5688.1940 & 2.104571 & $-$1.400 &---  & --- &--- & $-$1.400 &
& $-$0.30  & \phantom{+}0.00   & \phantom{+}0.00 & \phantom{+}0.00 \\
AlI     & 6696.185 & 4.021753 & $-$1.576 &$-$1.342 & $-$0.320 &$-$1.45$^1$ & $-$1.576 &
 & $-$0.30  & $-$0.20  & +0.10 & \phantom{+}0.00 \\
AlI     & 6696.788 & 4.021919 & $-$1.421 &$-$1.342  & $-$0.320 &$-$1.45$^1$ & $-$1.421 &
 & $-$0.30  & $-$0.20  & +0.10 & \phantom{+}0.00 \\
AlI     & 6696.788 & 4.021919 &$-$2.722 &$-$1.342  & $-$0.320 &$-$1.45$^1$ &$-$2.722 &
 & $-$0.30  & $-$0.20  & +0.10 & \phantom{+}0.00 \\
AlI     & 6698.673 & 3.142933 & $-$1.647 &---  & $-$0.320 &$-$1.78$^1$ & $-$1.647 &
& $-$0.30  & $-$0.20   &  \phantom{+}0.00 & \phantom{+}0.00 \\     
MgI     & 5528.4047 & 4.346096 & $-$0.620 &$-$0.498  & $-$0.620 &$-$0.25$^1$ & $-$0.498 &
& +0.40    & \phantom{+}0.00 & +0.15 & +0.20  \\
SiI     & 5665.555 & 4.920417 &$-$2.040 &-2.040  &$-$-2.04 &$-$1.94$^1$ &$-$2.04 &
& +0.20 & +0.15 & \phantom{+}0.00 & +0.15 \\
SiI     & 5666.690 & 5.616073 & $-$1.050 &---  & $-$1.795 &$-$1.74$^1$ & $-$1.74 &
& +0.30 & +0.15 & \phantom{+}0.00 & +0.15 \\
SiI     & 5690.425 & 4.929980 & $-$1.870 &$-$1.870  & $-$1.870 &$-$1.75$^1$ & $-$1.87 &
& +0.20 & +0.15 & +0.10 & +0.20 \\
SiI     & 5948.545 & 5.082689 & $-$1.230 &$-$1.231  & $-$1.230 &$-$1.17$^1$ & $-$1.30 &
& +0.30  & \phantom{+}0.00 & \phantom{+}0.00 & +0.30 \\
CaI     & 5601.277 & 2.525852 & $-$0.690 &$-$0.69  & $-$0.523 &$-$0.52$^1$ & $-$0.52 &
& \phantom{+}0.00  & $-$0.30 & +0.00 & +0.20 \\
CaI     & 5867.562 & 2.932710 & $-$0.801 &---  & $-$1.570 &$-$1.58$^1$ & $-$1.55 &
& \phantom{+}0.0  & \phantom{+}0.0 & +0.3 & +0.3 \\
CaI     & 6102.723 & 1.879467 & $-$0.890 &$-$0.79  & $-$0.793 &$-$0.65$^1$ & $-$0.793 &
& \phantom{+}0.00  & \phantom{+}0.00 & \phantom{+}0.00 & +0.05 \\
CaI     & 6122.217 & 1.885935 & $-$0.409 &$-$0.315  & $-$0.316 &$-$0.02$^1$ &  $-$0.20 &
& \phantom{+}0.00  & $-$0.30 & +0.30 & \phantom{+}0.00 \\
TiI     & 5689.459 & 2.296971 & $-$0.469 &$-$0.36  & $-$0.360 &$-$0.45$^1$ & $-$0.40 &
& \phantom{+}0.00  & \phantom{+}0.00 & \phantom{+}0.00 & \phantom{+}0.00 \\
TiI     & 5866.449 & 1.066626 & $-$0.840 &$-$0.840  & $-$0.840 &$-$0.71$^1$ & $-$0.84 &
& \phantom{+}0.0  & \phantom{+}0.0 & +0.15 & \phantom{+}0.00 \\
TiI     & 5922.108 & 1.046078 & $-$1.466 &$-$1.465  & $-$1.466 &$-$1.47$^1$ & $-$1.46 &
& \phantom{+}0.00  & \phantom{+}0.00 & \phantom{+}0.00 & \phantom{+}0.00 \\
TiI     & 5941.750 & 1.052997 & $-$1.510 &$-$1.52  & $-$1.51 &$-$1.54$^1$ & $-$1.53 &
& +0.15  & \phantom{+}0.00 & \phantom{+}0.00 & +0.10 \\
TiI     & 5965.825 & 1.879329 & $-$0.409 &$-$0.409  & $-$0.409 &$-$0.50$^1$ & $-$0.42 &
& \phantom{+}0.00  & +0.10 & +0.10 & +0.05 \\
TiI     & 5978.539 & 1.873295 & $-$0.496 &$-$0.496  & $-$0.496 &$-$0.51$^1$ & $-$0.53 &
& \phantom{+}0.0  & \phantom{+}0.0 & \phantom{+}0.0 & +0.2 \\
TiI     & 6064.623 & 1.046078 & $-$1.944 &$-$1.944  & $-$1.944 &$-$1.89$^1$ & $-$1.944 &
& \phantom{+}0.00  & $-$0.05 & \phantom{+}0.00 & \phantom{+}0.00 \\
TiI     & 6126.214 & 1.066626 & $-$1.425 &$-$1.424  & $-$1.425 &$-$1.33$^1$ & $-$1.425 &
& \phantom{+}0.00  & \phantom{+}0.00 & \phantom{+}0.00  & \phantom{+}0.00 \\
TiII     & 5154.0682 & 1.565869 & $-$1.920 &$-$1.92 & $-$1.750 &$-$1.60$^1$ & $-$1.75 &
& \phantom{+}0.00  & \phantom{+}0.00 & \phantom{+}0.00 & \phantom{+}0.00 \\
TiII     & 5336.771 & 1.581911 & $-$1.700 &$-$1.70  & $-$1.59 &$-$1.54$^1$,$-$1.60$^2$ & $-$1.70 &
& \phantom{+}0.00  & \phantom{+}0.00 & \phantom{+}0.00 & \phantom{+}0.00 \\
TiII     & 5381.0212 & 1.565869 &$-$-2.080 &$-$1.70  & $-$1.92 &$-$1.95$^1$,$-$1.97$^2$ &$-$2.08 &
& \phantom{+}0.00  & \phantom{+}0.00 & \phantom{+}0.00 & \phantom{+}0.00 \\
TiII     & 5418.751 & 1.581911 & $-$1.999 &-2.002  &$-$-2.00 &-2.10$^1$,-2.13$^2$ &$-$2.13 &
& +0.30  & +0.30 & +0.30 & +0.10 \\
\noalign{\vskip 0.1cm}
\noalign{\hrule\vskip 0.1cm}
\noalign{\vskip 0.1cm}  
\hline                  
\end{tabular}
\end{flushleft}
\end{table*}

\begin{table*}
\caption[1]{Mean abundances of C, N, odd-Z elements Na, Al,
and $\alpha$-elements O, Mg, Si, Ca, Ti. For
Na and Al the abundance ratios in parenthesis are corrected
for a trend with effective temperature. }
\begin{flushleft}
\tabcolsep 0.15cm
\begin{tabular}{cccccccccccc}
\noalign{\smallskip}
\hline
\noalign{\smallskip}
\hline
\noalign{\smallskip}
{\rm star} & [C/Fe] & [N/Fe] & [Na/Fe] & [Al/Fe] & [O/Fe] & [Mg/Fe] & [Si/Fe] & [Ca/Fe] & [TiI/Fe] &[TiII/Fe] &  \cr
\noalign{\vskip 0.2cm}
\noalign{\hrule\vskip 0.2cm}
\noalign{\vskip 0.2cm}
B-107 & \phantom{+}0.00  & ---& $-$0.22(+0.03) & $-$0.30(+0.28) & +0.50 & +0.33 & +0.17 & +0.16 & +0.03 & +0.17 \cr
B-122 & $-$0.20 & +0.70& $-$0.04(+0.09) & $-$0.20(+0.18) & +0.20 & +0.10  & +0.06 & +0.00 & +0.03  & +0.15 \cr
B-128 & \phantom{+}0.10  & +0.60 & +0.01(+0.01) & +0.08(+0.08) & +0.23 & +0.23  & +0.14 & +0.20 & +0.05  & +0.17  \cr
B-130 & \phantom{+}0.00   & +0.70 & $-$0.01(+0.05) &  \phantom{+}0.00(+0.26) & +0.50 & +0.27  & +0.13 & +0.15 & +0.03  & +0.18 \cr
\noalign{\vskip 0.2cm}
\noalign{\hrule\vskip 0.2cm}
\noalign{\vskip 0.2cm}
Mean & $-$0.03   & +0.67 & $-$0.07(+0.05) &  $-$0.11(+0.20) & +0.36 & +0.23  & +0.13 & +0.13 & +0.04  & +0.17 \cr
\noalign{\smallskip} \hline \end{tabular}
\label{alpha}
\end{flushleft}
\end{table*}


\subsection{Heavy elements}
\label{Sect:heavy}

We derive the abundances of the neutron-capture elements 
Sr, Y, Zr, La, and Ba and the reference r-element Eu.

For the heavy elements Sr, Y, Zr, La, Ba, and Eu, 
the first ionization stages dominate 
the total abundances in the studied effective temperature range 
(4000 $<$ T$_{\rm eff}$ $<$ 6000 K), therefore
we preferentially used lines of ionized species. Because of 
  a lack of reliable ionized lines, we also measured abundances
from lines of \ion{Sr}{I}, \ion{Y}{I}, and \ion{Zr}{I}.

We checked the solar and Arcturus spectra for lines of these elements
by computing synthetic spectra and verifying the change in their 
intensities that is due to changes in abundances. The lines were selected 
when their contribution to the line in question was dominant,
as well as  by checking which of them were present and measurable in
the sample stars, even if in some cases they are too faint 
in the solar and/or Arcturus spectra.

The wavelengths, excitation potentials, and oscillator strengths
were gathered from the line lists of the Kur\'ucz (1993) 
websites\footnote{http://www.cfa.harvard.edu/amp/amp\-data/ku\-rucz23/\-se\-kur.\-html}
\footnote{http://kurucz.harvard.edu/atoms.html}, the
National Institute of Standards \& Technology (NIST, Martin et al. 2002)
\footnote{http://physics.nist.gov/PhysRefData/ASD/lines$_-$form.html}, 
and VALD (Piskunov et al. 1995).

The hyperfine structure (HFS) for the studied lines of 
\ion{La}{II}, \ion{Ba}{II}, and \ion{Eu}{II} were taken into account.
We computed the splitting of lines
by employing a code made available by Andrew McWilliam, following the
calculations described by Prochaska \& McWilliam (2000). 

{\it Barium:} The nuclear spin is I=1.5 and the nuclides
 $^{138}$Ba and $^{137}$Ba, Ba$^{136}$, Ba$^{135}$ and Ba$^{134}$ 
contribute with 71.7\%, 11.23\%,
 7.85\%, 6.59\%, and 2.42\%  to the total abundance respectively
(Lodders 2009).
Experimental data on hyperfine coupling constants, 
the magnetic dipole A-factor,
and the electric quadrupole B-factor
were adopted from Biehl (1976) and Rutten (1978),
as given in Table A.1.
We computed the HFS for Ba.
 The resulting line lists with the HFS
 for the \ion{Ba}{II} 6141.713 and
6496.897 {\rm \AA} lines are given in Table A.2.


{\it Lanthanum:} 
 The nuclear spin  of the nuclide  $^{139}$La that
contributes with 99.911\% to the lanthanum abundance is I=7/2.
Hyperfine coupling constants A and B 
were adopted from Biehl (1976) and Lawler et al. (2001a)
 as given in Table A.3.
For the LaII 6262.287 {\rm \AA} line we adopted the HFS calculation
by Van der Swaelmen (2013).

{\it Europium:}
 The nuclear spin  of the nuclides $^{151}$Eu,
and $^{153}$Eu is I=5/2, and their isotopic proportions 
 are 47.8\% and 52.2\% (Lawler et al. 2001b). 
Hyperfine coupling constants A and B are taken from
 Lawler et al. (2001b).
The HFS splitting was adopted from line lists
by Hill et al. (2002), and show a perfect fit
to the solar and Arcturus lines.
The EuII 6173.029 {\rm \AA} line is too faint in all sample stars,
and could not be measured.

For Sr, Y and Zr no HFS constants were found in the literature.
We comment below on lines of Y and Sr used in C11:

{\it Yttrium:}
The \ion{Y}{I} and  \ion{Y}{II} lines employed are reported in Table \ref{lines2}.
The main Y abundance indicator used in C11 was the 
\ion{Y}{II} 6613.733 {\rm \AA}
line, and we proceed with a detailed description of this line. 
The \ion{Y}{II} 6613.733 line has a blend with the \ion{Fe}{I}
6613.830 {\rm \AA} line, which is  0.097 {\rm \AA} apart.
The log gf values of these two lines
 given in the literature are too strong, and to be able
to use this line, we fitted astrophysical log gf values. 
The solar line was fitted to
the  solar spectrum observed with UVES, and an oscillator strength
of log gf = -1.2 was chosen. The blending \ion{Fe}{I} 6613.830 {\rm \AA}
was fitted with log gf = -5.8. The list of log gf values around this
line is given in Table A.4.

\begin{figure}
\centering
\psfig{file=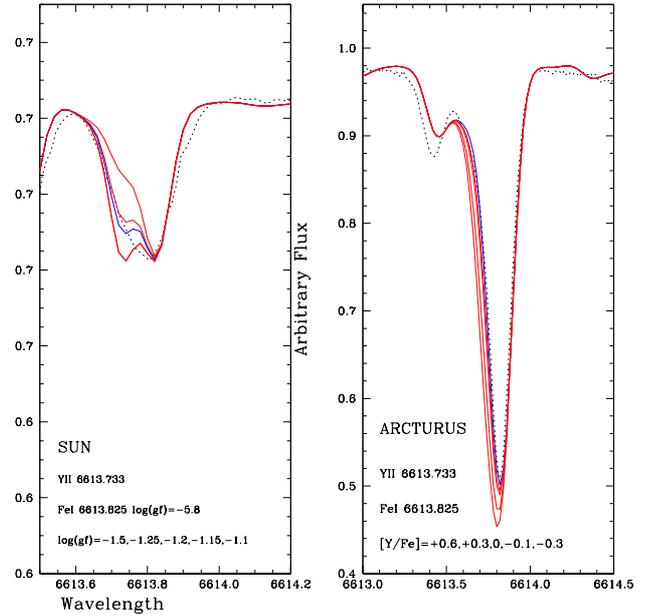,angle=0.,width=9.0 cm}
\caption{\ion{Y}{II} 6613.733 fittings on the solar and Arcturus spectra. 
Observed spectra ({\it dashed lines}); synthetic
spectra ({\it red solid lines}); best fit ({\it blue solid line}).}
\label{y6613solarc} 
\end{figure}

\begin{figure}
\centering
\psfig{file=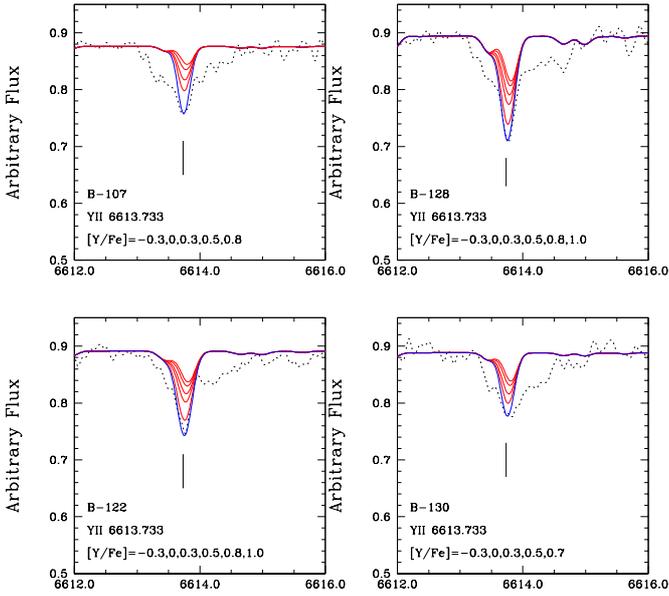,angle=0.,width=9.0 cm}
\caption{\ion{Y}{II} 6613.733 {\rm \AA} line in the four sample stars,
compared with their new UVES spectra. Dotted black line: UVES spectrum;
red solid lines: synthetic spectra computed with different [Y/Fe] values as indicated in the panels; blue solid lines correspond to best fits.}  
\label{yuves} 
\end{figure}

 The fitting was then applied to the spectrum
of Arcturus using these log gf values, and 
was computed with different abundances
of yttrium: the [Y/Fe]=-0.3 value assumed from the other Y lines fits
the line well,
and higher Y abundances will show as a stronger line, whereas a lower Y cannot be seen,
only a lower limit is possible in this case. These fits are shown in Fig. 
\ref{y6613solarc}. On the left wing of the blend \ion{Y}{II} 6613.733 + \ion{Fe}{I} 6613.830
{\rm \AA} line, there are two additional \ion{Ti}{I} 
 6613.599 and 6613.620 {\rm \AA} lines
in the VALD data base that are not given in the Kur\'ucz line list,
with only the   \ion{Ti}{I} 6613.626 {\rm \AA} given in both data bases.
 The log gf of these three \ion{Ti}{I} lines were corrected by
-0.3 dex, as given in Table \ref{srblends} to fit 
the Arcturus spectrum.
The fits to the new UVES spectra are shown in Fig. \ref{yuves}.
 Even though this line
gives higher Y abundances than the other lines, 
no suitable Y lines are available, therefore the
results from this line were taken into account together with the others.
Previous GIRAFFE spectra with lower resolution were clearly 
 better fitted with even
higher Y abundances. 

\begin{figure}
\centering
\psfig{file=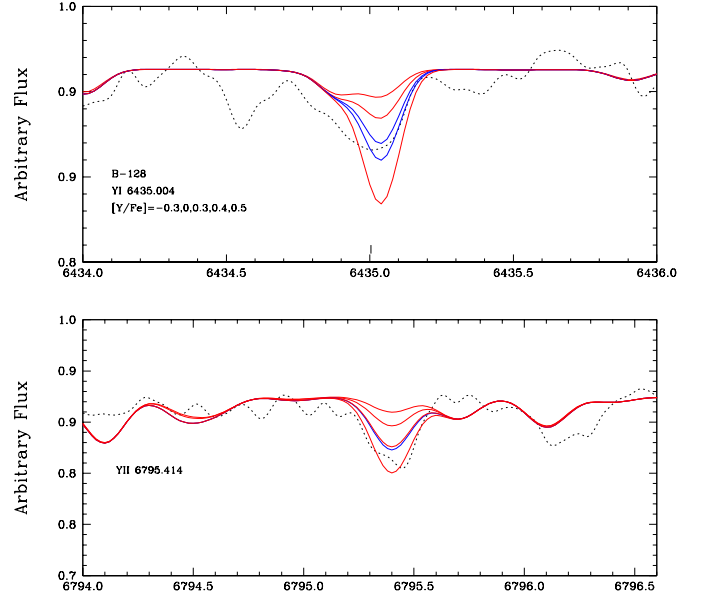,angle=0.,width=9.0 cm}
\caption{\ion{Y}{I} 6435 and\ion{Y}{I} 6795 {\rm \AA} lines
 in B-128.  Dotted black line: UVES spectrum;
red solid lines: synthetic spectra computed with different [Y/Fe] values as indicated in the panels; blue solid line: best fit
for the YI 6435 line the [Y/Fe] = +0.3 and +0.4 are blue,
and [Y/Fe] = +0.35 was adopted. 
}
\label{y} 
\end{figure}

The fits to \ion{Y}{I} 6435 and \ion{Y}{I} 6795 {\rm \AA} lines
 in star B-128 are shown in Fig. \ref{y}.

 In C11 stronger Y abundances were derived, because of a
slightly lower log gf value of the blending
 \ion{Fe}{I}  6613.830 {\rm \AA} line with log gf = -5.85,
instead of -5.80, and 
 the  \ion{Y}{II} 6613.733 line with a log gf = -1.25 instead of -1.20.
 These lower log gf values, combined to the lower resolution of the
spectra, led to higher Y abundances.

{\it Strontium:}
The \ion{Sr}{I} 6550.244 {\rm \AA} line is blended with nine lines, 
 four of which coincide with the \ion{Sr}{I} line
(Table \ref{srblends}), and the log gf values
of these lines, currently listed in the VALD data base, 
are stronger than the observed lines for the Sun,
Arcturus, and the sample stars. We conclude that some of 
their log gf values are clearly inaccurate,
and since they are blended with each other, it is not possible 
to derive astrophysical log gf
relative values with confidence. 
This line was therefore discarded in the present work.

In C11 we used a line list with astrophysically fitted
 log gf values, and the lower resolution of
 the spectra appeared to show a strong SrI 6550 line. 
Fig. \ref{srcomp} shows
a comparison of the previous and current spectra for star B-128.
 The Sr abundances presented in C11, based on the
\ion{Sr}{I} 6550.244 {\rm \AA}, should be disconsidered.

\begin{figure}
\centering
\psfig{file=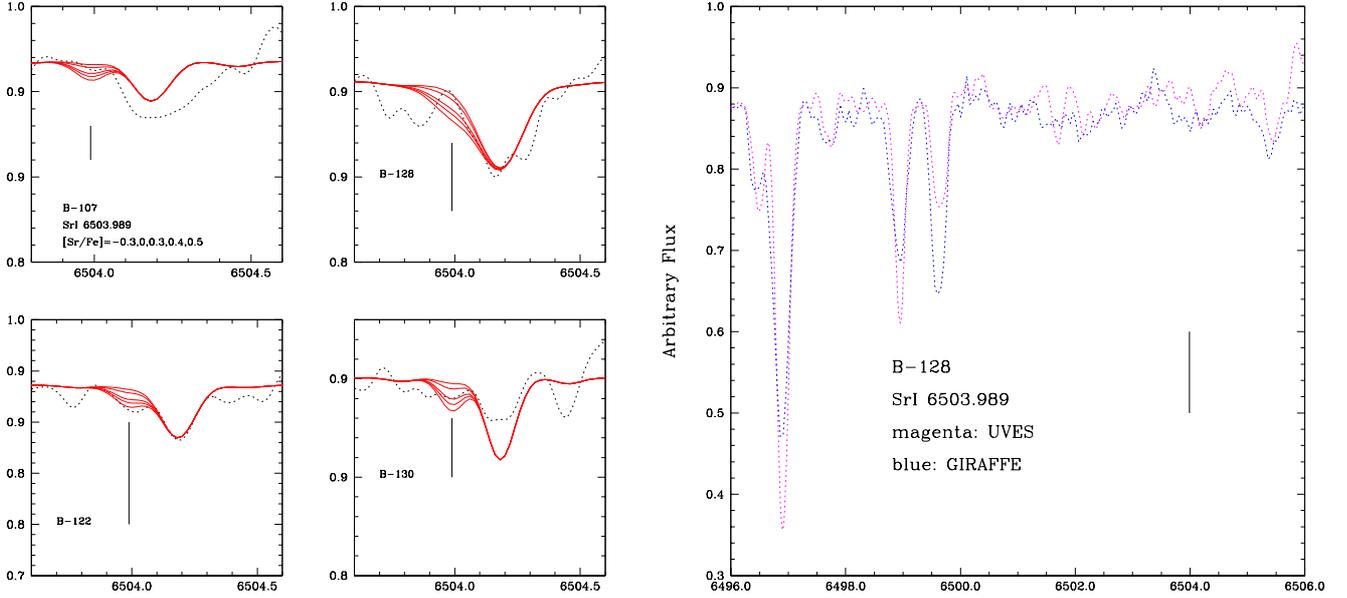,angle=0.,width=9.0 cm}
\caption{\ion{Sr}{I} 6503.989 {\rm \AA} line
 in the four sample stars. Symbols: dotted black line: UVES
spectrum; red solid lines: synthetic spectra computed with
different [Sr/Fe] ratios. }
\label{sr} 
\end{figure}

\begin{figure}
\centering
\psfig{file=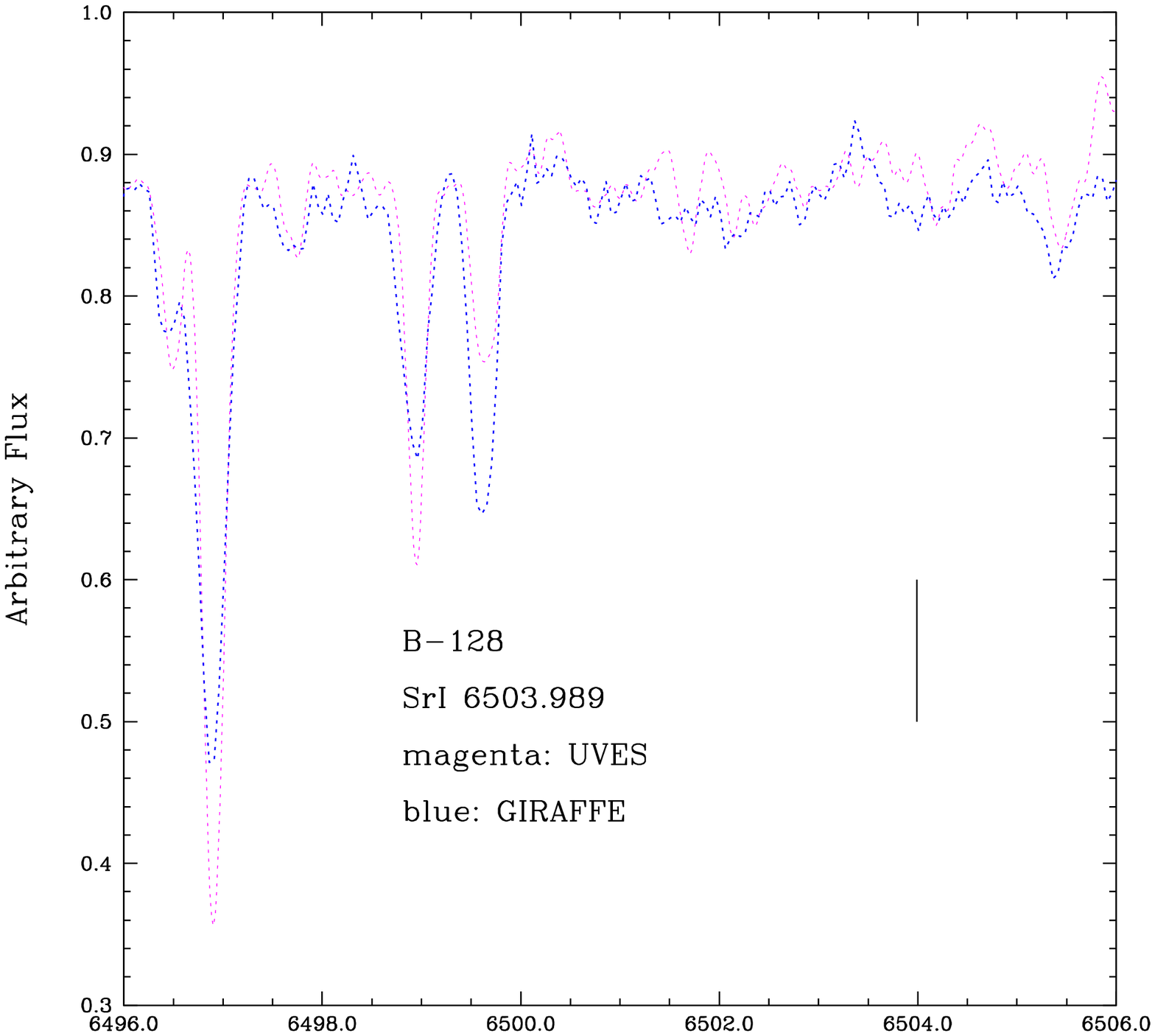,angle=0.,width=9.0 cm}
\psfig{file=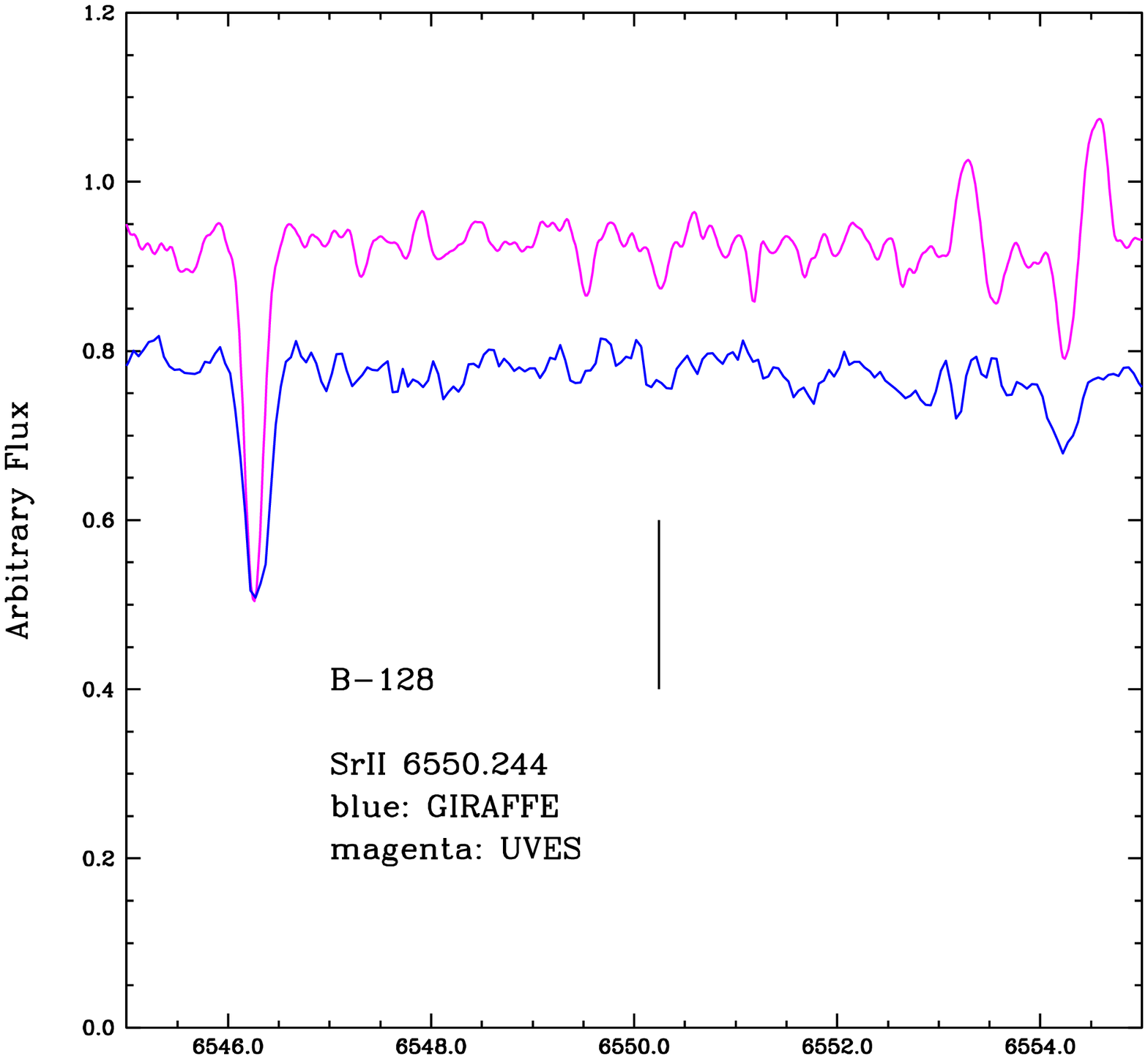,angle=0.,width=9.0 cm}
\caption{\ion{Sr}{I} 6503.989 and \ion{Sr}{I} 6550.244 {\rm \AA} lines
 in the same star B-128, shown in spectra of UVES
from the present work, observed in 2011-2012,
and GIRAFFE spectrum studied in B09. }
\label{srcomp} 
\end{figure}

\ion{Sr}{I} 6503.989 {\rm \AA} instead is relatively unblended,
 showing a strong line on the red side, for which we 
derived astrophysical log gf values (since the VALD log gf 
values are again too strong), and blending lines are indicated in 
Table \ref{srblends}.
 This line is the main indicator of the Sr abundance in the present work.
Fig. \ref{sr} shows the \ion{Sr}{I} 6503.989 {\rm \AA} line
 in the four sample stars.
There are very few derivations of Sr abundances in bulge stars
 in the literature, given that the suitable lines are located in the blue
part of the spectrum, while in the visible region the lines are
faint or blended. 

\begin{figure}
\centering
\psfig{file=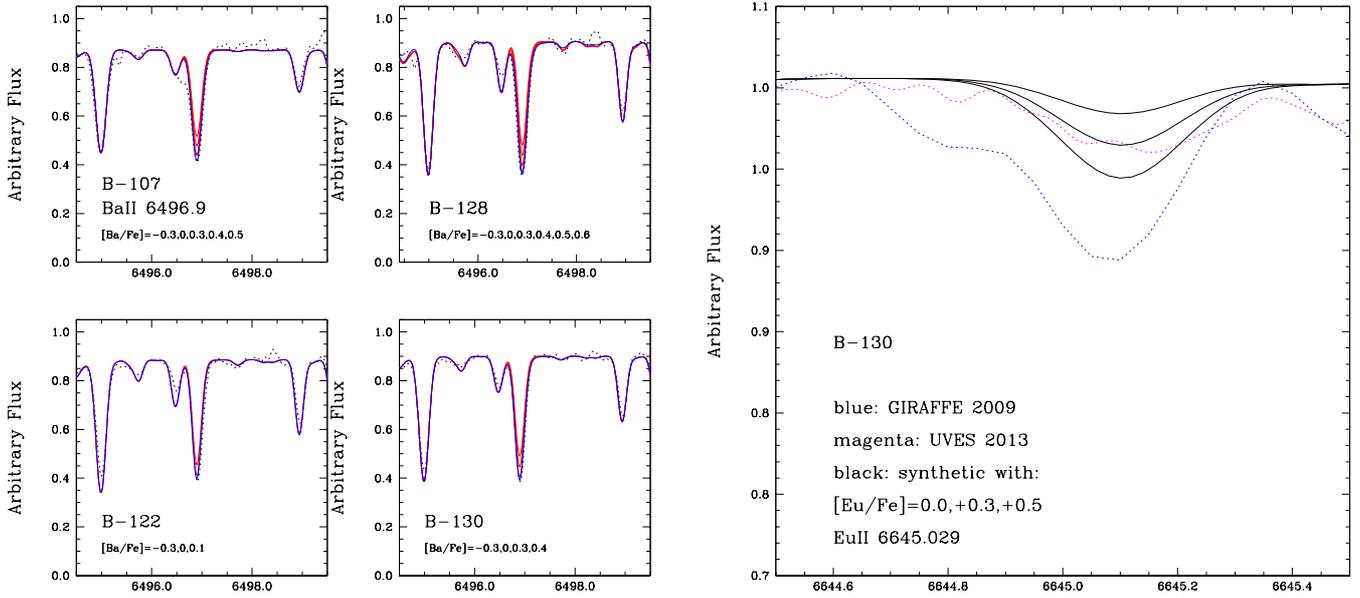,angle=0.,width=9.0 cm}
\caption{\ion{Ba}{II} 6946 {\rm \AA} line in the four sample stars.
 Red lines: synthetic spectra computed with several [Ba/Fe] values as
indicated in the panels; blue lines: best-fit synthetic spectra.} 
\label{ba} 
\end{figure}

\begin{figure}
\centering
\psfig{file=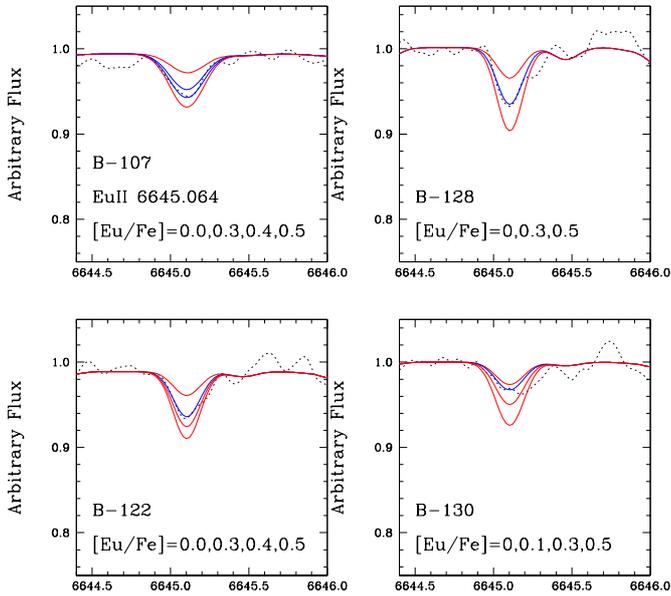,angle=0.,width=9.0 cm}
\caption{\ion{Eu}{II} 6645 {\rm \AA} line in the four sample stars.
Symbols are the same as in Fig. \ref{ba}. }
\label{eu} 
\end{figure}

 Figures \ref{ba} and \ref{eu} show the
\ion{Ba}{II} 6946 {\rm \AA} and \ion{Eu}{II} 6645 {\rm \AA} 
 lines in the four sample stars.

\begin{figure}
\centering
\psfig{file=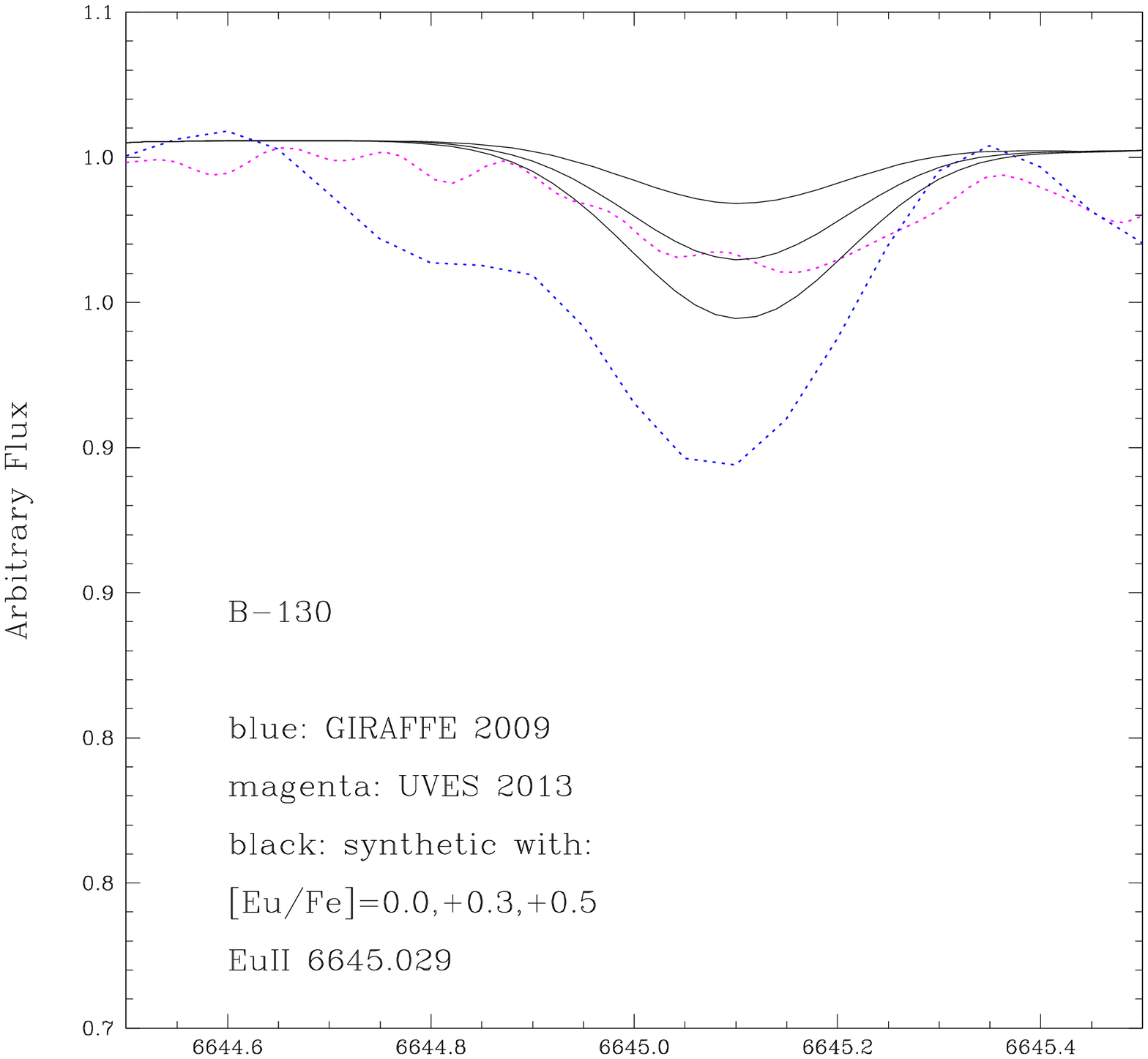,angle=0.,width=9.0 cm}
\psfig{file=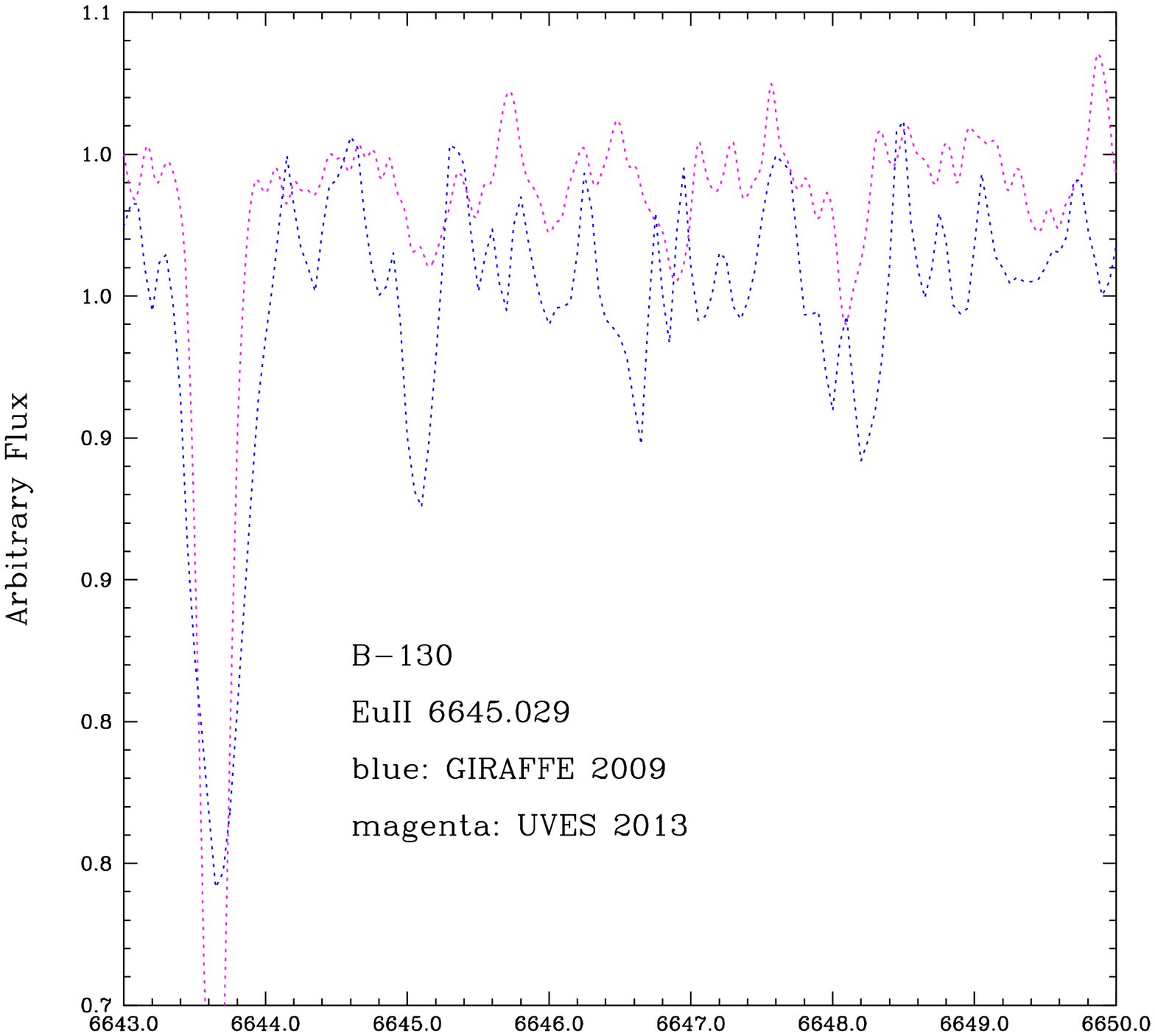,angle=0.,width=9.0 cm}
\caption{\ion{Eu}{II} 6645 {\rm \AA} line
 in the same star B-130, shown in spectra of UVES
from the present work, observed in 2011-2012,
and in the GIRAFFE spectrum studied in B09.
Upper panel: \ion{Eu}{II} 6645 {\rm \AA} line; 
lower panel: the same spectral
 region enlarged. The magenta dotted line represents the UVES
spectrum, the blue dotted line the GIRAFFE spectrum, and
the blue solid lines plot the synthetic
spectra computed with [Eu/Fe]=0, +0.3, +0.5. }
\label{eu2} 
\end{figure}

 To make the differences in
Eu abundance from B09, C11, and the present work understandable,
we show in Fig. \ref{eu2} the spectra of star B-130 in the
region of the \ion{Eu}{II} 6645 {\rm \AA} line
 obtained with UVES in the present work, observed in 2011-2012,
and the HR15 setup of the GIRAFFE spectrum studied in B09.
The much stronger noise in the GIRAFFE spectrum with respect
to the present UVES spectrum is clearly illustrated,
showing that the spectra of the HR15 setup have 
a lower S/N than previously reported.

In conclusion, given the large uncertainties in the Sr abundances,
 based on a unique
 and faint line, we here focus our discussion on Y and Ba, 
similar to the decision made in C11. 

\begin{table*}
\begin{flushleft}
\caption{Central wavelengths from NIST or Kur\'ucz line lists and total oscillator strengths
from line lists by Kur\'ucz, NIST, and VALD, the literature, and adopted
values. In Col. 7, literature oscillator strength values are taken from the
references: 1 Hannaford et al. 1982, 2 Lawler et al. (2001a),
 3 Lawler et al. (2001b),
4 Jacobson \& Friel (2013), and 5 van der Swaelmen (2013). 
 Line-by-line abundances of heavy elements for the four sample stars are given.}
\label{lines2}      
\centering 
\small         
\begin{tabular}{l@{}c@{}c@{}c@{}c@{}c@{}c@{}c@{}c@{}ccccccccc}     
\noalign{\smallskip}
\hline\hline    
\noalign{\smallskip}
\noalign{\vskip 0.1cm} 
species & {\rm $\lambda$} ({\rm \AA}) & \phantom{-}{\rm $\chi_{ex}$ (eV)} & 
\phantom{-}{\rm gf$_{Kurucz}$} &
 \phantom{-}{\rm gf$_{NIST}$} & \phantom{-}{\rm gf$_{VALD}$} &
 \phantom{-}{\rm gf$_{literature}$} & \phantom{-}{\rm gf$_{adopted}$}&
& \phantom{-}\phantom{-}\hbox{B-107} &\hbox{B-122} &\hbox{B-128} &\hbox{B-130}   \\
\noalign{\vskip 0.1cm}
\noalign{\hrule\vskip 0.1cm}
\noalign{\vskip 0.1cm}
EuII     & 6437.640 & 1.319712 & $-$0.276 &---  & $-$0.320 &$-$0.32$^3$ & $-$0.32 &
& +0.45  & +0.30   & +0.30 & +0.30 \\
EuII     & 6645.064 & 1.379816 & +0.204 &---  & +0.120 &+0.12$^3$ & +0.12 &
& +0.40  & +0.30   & +0.30 & +0.10 \\
BaII & 6141.713 & 0.703636 & $-$0.076 & \phantom{+}0.0  &$-$0.076 & $-$0.076 & \phantom{+}0.00 &
& +0.40   & \phantom{+}0.0   & +0.50  & +0.10 \\
FeI     & 6141.730 & 3.602 & $-$1.60 &---  & $-$1.459 &--- & $-$1.60 &
& --- & --- & --- & ---  \\
BaII & 6496.897 & 0.604321 & $-$0.377 &$-$0.407  &$-$0.32 & --- & $-$0.32 &
& +0.50   & +0.10  & +0.60  & +0.35   \\
LaII & 6262.287  & 0.403019 & $-$1.240 & --- & $-$1.220 & $-$1.22,$-$1.60$^{5}$ & $-$1.60 &
& \phantom{+}0.00   & +0.40 & +0.40 & \phantom{+}0.00   \\
LaII & 6320.376  & 0.172903 & $-$1.610 & --- & $-$1.562 &-- & $-$1.56  &
& +0.30   & +0.30   & +0.30 & +0.10 \\
LaII & 6390.477  & 0.321339 & $-$1.450 & --- & $-$1.410 &$-$1.41 & $-$1.41  &
& +0.30   & +0.30   & +0.40 & \phantom{+}0.00  \\
YII & 5473.388    & 1.738160 & $-$1.02   & $-$1.01  & $-$1.020 & $-$1.02$^1$ & $-$1.02  &
 & +0.30   & +0.15   & +0.30 & +0.10 \\
YII & 5544.611    & 1.738160 & $-$1.09   & $-$1.08  & $-$1.090 & $-$1.09$^1$ & $-$1.09 &
& +0.20   & \phantom{-}0.00    & +0.30 & \phantom{+}0.00 \\
YII &  5546.009   & 1.748055 & $-$1.10   & $-$1.10  & $-$1.100 & $-$1.10$^1$ & $-$1.10  &
& +0.30   & +0.30   & +0.30 & --- \\
YI & 6435.004     & \phantom{+}0.065760  & $-$0.820  & $-$0.83 & $-$0.820 & $-$0.82$^1$,$-$1.07$^5$ & $-$0.82 &
& +0.30 & \phantom{+}0.00 &+0.35 &--- \\
YII & 6613.733    & 1.011123 & $-$6.689   & ---  & $-$5.587 & --- & $-$5.80 &
& +0.80 & +1.00 & +1.00  & +0.70 \\
FeI & 6613.825    & 1.748055 & $-$1.110   &$-$1.110  & $-$1.110 & $-$1.11$^1$& $-$1.15 &
& --- & --- & ---  & --- \\
YII & 6795.414    & 1.738160 & $-$1.190   & ---   & $-$1.190 & ---& $-$1.19 &
& \phantom{+}0.00
   &  \phantom{+}0.00   & +0.30   & +0.10 \\
ZrII & 5112.270   & 1.665034 & $-$0.590  & ---   & $-$0.850 & --- & $-$0.59 &
& \phantom{+}0.00   & +0.10   & +0.40   & \phantom{+}0.00 \\  
ZrII & 5350.090   & 1.826828 & $-$1.240  & ---   &$-$1.240 & --- & $-$1.24  &
& ---    & +0.30   & ---    & ---  \\
ZrII & 5350.350   & 1.772925 & $-$1.276  & ---   &$-$1.160 & --- & $-$1.16 &
 & +0.30  & +0.30   & ---    & ---     \\
ZrI  & 6127.475   & 0.153855 & $-$1.06 & ---   & $-$1.060 & $-$1.18$^4$,$-$1.05$^5$ & $-$1.18 &
 & +0.30    &   \phantom{+}0.00    & +0.40  & \phantom{+}0.00  \\
ZrI  & 6134.585   & \phantom{+}0.00    & $-$1.28 & ---   & $-$1.280 & $-$1.426$^4$,$-$1.28$^5$ & $-$1.43 &
& ---      &  \phantom{+}0.00    & +0.35 & ---  \\
ZrI  & 6143.252   & \phantom{+}0.070727 & $-$1.10 & ---   & $-$1.100 & $-$1.252$^4$,$-$1.10$^5$ & $-$1.50 &
& ---     &  +0.15    & +0.30  & \phantom{+}0.00 \\
SrI & 6503.989 & 2.258995   & +0.26   & $-$0.05 &  +0.320 & ---  & $-$0.05 &
 & \phantom{+}0.00   & +0.40  & \phantom{+}0.00  & +0.30  \\ 
SrI & 6791.016 & 1.775266   & $-$0.720   & $-$0.73  &  $-$0.73 & --- & $-$0.73 &
 & ---   &  ---    & +0.40 & --- \\
\noalign{\vskip 0.1cm}
\noalign{\hrule\vskip 0.1cm}
\noalign{\vskip 0.1cm}  
\hline                  
\end{tabular}
\end{flushleft}
\end{table*}

\begin{table*}
\caption[1]{Mean heavy element abundances, compared with results
 from B09 and C11. }
\begin{flushleft}
\small
\tabcolsep 0.15cm
\begin{tabular}{ccccccccccccccccccc}
\noalign{\smallskip}
\hline
\noalign{\smallskip}
\hline
\noalign{\smallskip}
{\rm star} & [Eu/Fe] & [Eu/Fe] & [Ba/Fe] & [Ba/Fe] & [La/Fe] & [La/Fe] & [Y/Fe] & [Y/Fe] & [Sr/Fe] &
[Sr/Fe] & [Zr/Fe] &  [Ba/Eu] &  [Y/Ba] &  [Sr/Ba] &  \cr
\noalign{\smallskip}
 & old & new & old & new & old & new & old & new & old &
new & new & { new} & { new} & { new} \cr
\noalign{\vskip 0.2cm}
\noalign{\hrule\vskip 0.2cm}
\noalign{\vskip 0.2cm}
B-107 & \phantom{+}0.00  & +0.40 & +0.50 & +0.45 & +0.50 & +0.20  & +1.00 & +0.32 & +1.30  & \phantom{+}0.00 & +0.20 & +0.05 & $-$0.13 &$-$0.45 &  \cr
B-122 & +0.30 & +0.30 & +0.60 & +0.05 & +0.30 & +0.35  & +1.20 & +0.24 & +0.50  & +0.40 & +0.10 & $-$0.25  & +0.19 & +0.35 &  \cr
B-128 & \phantom{+}0.00  & +0.30 & +0.90 & +0.55 & ---    & +0.35  & +1.50 & +0.43 & +1.50 & +0.20 & +0.40 & +0.25  &$-$0.12 & $-$0.35 &  \cr
B-130 & +0.80 & +0.20 & +0.25 & +0.22 & ---    & \phantom{+}0.00    & +1.20 & +0.23 & ---   & +0.30 & \phantom{+}0.00 & +0.02 &+0.01 & $+$0.08 &  \cr
\noalign{\vskip 0.05cm}
\noalign{\hrule\vskip 0.05cm}
Mean & --- & +0.30 & --- & +0.32 & ---  & +0.23  & --- & +0.31 & --- & +0.23 & +0.18 & +0.02 &$-$0.01 & $-$0.14 &  \cr
\noalign{\smallskip} \hline \end{tabular}
\label{heavy}
\end{flushleft}
\end{table*}

\subsection{Errors}

 The errors due to uncertainties in spectroscopic parameters are given 
in Table \ref{errors}, applied to the sample star NGC 6522: B-128.
The error  on the slope in the  FeI $vs.$ ionization potential implies
an error in  the temperature of $\pm$100 K   for the sample  
stars. An uncertainty of the order of  0.2 km~s$^{-1}$ on the microturbulence
velocity is estimated from the imposition of constant value of [Fe/H] as
a function of EWs.
 Errors are given on FeI and FeII abundances,
  and other element abundance ratios, induced  by   a change of  $\Delta$T${\rm
eff}$=+100 K, $\Delta$log  g  =+0.2,  $\Delta$v$_{\rm  t}$ =  0.2  km~s$^{-1}$, 
and a total error estimate is given in the last column of Table \ref{errors}.
 Additionally, an uncertainty of 0.8 m{\rm \AA}
in EWs of Fe lines yields a metallicity uncertainty of $<$0.02 dex.

The errors on the abundance ratios [X/Fe] were computed by fitting the lines
with the modified atmospheric model. 
The error given is the abundance difference needed
to reach the final abundances reported.

\begin{table}
\caption{Abundance uncertainties for star B-128,
 for uncertainties of $\Delta$T$_{\rm eff}$ = 100 K,
$\Delta$log g = 0.2, $\Delta$v$_{\rm t}$ = 0.2 km s$^{-1}$ and
corresponding total error. The errors are to be
added to reach the reported abundances.} 
\label{errors}
\begin{flushleft}
\small
\tabcolsep 0.15cm
\begin{tabular}{lcccc}
\noalign{\smallskip}
\hline
\noalign{\smallskip}
\hline
\noalign{\smallskip}
\hbox{Abundance} & \hbox{$\Delta$T} & \hbox{$\Delta$log $g$} & 
\phantom{-}\hbox{$\Delta$v$_{t}$} & \phantom{-}\hbox{($\sum$x$^{2}$)$^{1/2}$} \\
\hbox{} & \hbox{100 K} & \hbox{0.2 dex} & \hbox{0.2 kms$^{-1}$} & \\
\hbox{(1)} & \hbox{(2)} & \hbox{(3)} & \hbox{(4)} & \hbox{(5)} \\
\noalign{\smallskip}
\hline
\noalign{\smallskip}
\noalign{\hrule\vskip 0.1cm}
\hbox{[FeI/H]}       &  +0.06          &\phantom{+}0.00  & $-$0.08         &\phantom{+}0.10 \\
\hbox{[FeII/H]}      &  $-$0.07        &  +0.11          & $-$0.07         &\phantom{+}0.14  \\
\hbox{[C/Fe]}        &  +0.02          &  +0.02   & \phantom{+}0.00        &\phantom{+}0.03  \\
\hbox{[N/Fe]}        &  +0.15          &  +0.10   & \phantom{+}0.00        &\phantom{+}0.18  \\
\hbox{[O/Fe]}        &  +0.06          &\phantom{+}0.10  &\phantom{+}0.00  &\phantom{+}0.12  \\
\hbox{[NaI/Fe]}      &\phantom{+}0.00  &\phantom{+}0.00  &\phantom{+}0.00   &\phantom{+}0.00  \\
\hbox{[Al/Fe]}       &  +0.05          &\phantom{+}0.00   &\phantom{+}0.00  &\phantom{+}0.05  \\
\hbox{[MgI/Fe]}      &  +0.03          &  +0.01          &\phantom{+}0.00   &\phantom{+}0.02 \\
\hbox{[SiI/Fe] }     & +0.01           &  +0.03          &\phantom{+}0.00   &\phantom{+}0.02  \\
\hbox{[CaI/Fe]}      &  +0.02          &  $-$0.02          &\phantom{+}0.00   &\phantom{+}0.02  \\
\hbox{[TiI/Fe]}      &  +0.10          &  $-$0.01          &\phantom{+}0.00   &\phantom{+}0.10  \\
\hbox{[TiII/Fe]}     &\phantom{+}0.00  &  +0.08          & $-$0.02          &\phantom{+}0.10  \\
\hbox{[SrI/Fe]}      &$-$0.02          &\phantom{+}0.00  &\phantom{+}0.00   &\phantom{+}0.02  \\
\hbox{[YI/Fe]}       &  +0.15          & +0.15           & \phantom{+}0.00  &\phantom{+}0.21  \\
\hbox{[YII/Fe]}      &  +0.20          & +0.15           & \phantom{+}0.00  &\phantom{+}0.25 \\
\hbox{[ZrI/Fe]}      &  +0.20          & $-$0.01           & \phantom{+}0.00  &\phantom{+}0.20  \\
\hbox{[BaII/Fe]}     &  +0.10          & +0.15           & $-$0.15          &\phantom{+}0.23  \\
\hbox{[LaII/Fe]}     &  +0.05          & +0.15           &\phantom{+}0.00   &\phantom{+}0.16  \\
\hbox{[EuII/Fe]}     &\phantom{+}0.00  & +0.12           &\phantom{+}0.00   &\phantom{+}0.12  \\
\noalign{\smallskip} 
\hline 
\end{tabular}
\end{flushleft}
\end{table}


\section{Discussion}
We  derived  a mean metallicity of [Fe/H]=$-$0.95$\pm$0.15, 
which agrees well the results of B09.
The final mean abundances are given in Table
\ref{alpha} for C, N, odd-Z elements Na, Al, and the $\alpha$-elements   
and in Table \ref{heavy} for the  heavy elements.

To better place the derived abundances
in the context of bulge studies, we compare the present results with literature
abundances of heavy elements in bulge stars.

Johnson et al. (2012) derived abundances of the heavy elements  Zr, La, and Eu
in common with our element abundance derivation. Their observations were of
red giants in Plaut's field, located at l=0$^\circ$, b=-8$^{\circ}$, and 
 l=-1$^\circ$, b=-$8\fdg5$, and their stars have metallicities
in the range $-$1.6 $\leq$ [Fe/H] $\leq$ +0.5. 

Yong et al. (2014) studied seven stars in the globular cluster M62 (NGC 6266), located at  J(2000)
17$^{\rm h}$01$^{\rm h}$$12\farcs80$,  $-$30$^{\circ}$06'$49\farcs4$, 
l = $353\fdg57$, 
b = $7\fdg32$, which is within the bulge volume.   

Bensby et al. (2013 and references therein) presented
 element abundances of 58 microlensed bulge
dwarfs and subgiants of the Galactic bulge. Their study included
 the abundances of the
heavy elements Y and Ba. Because the authors report ages,
 including younger stellar populations
present in the bulge, we selected a subsample of stars 
with ages older than 11 Gyr.

\begin{figure*}
\centering
\psfig{file=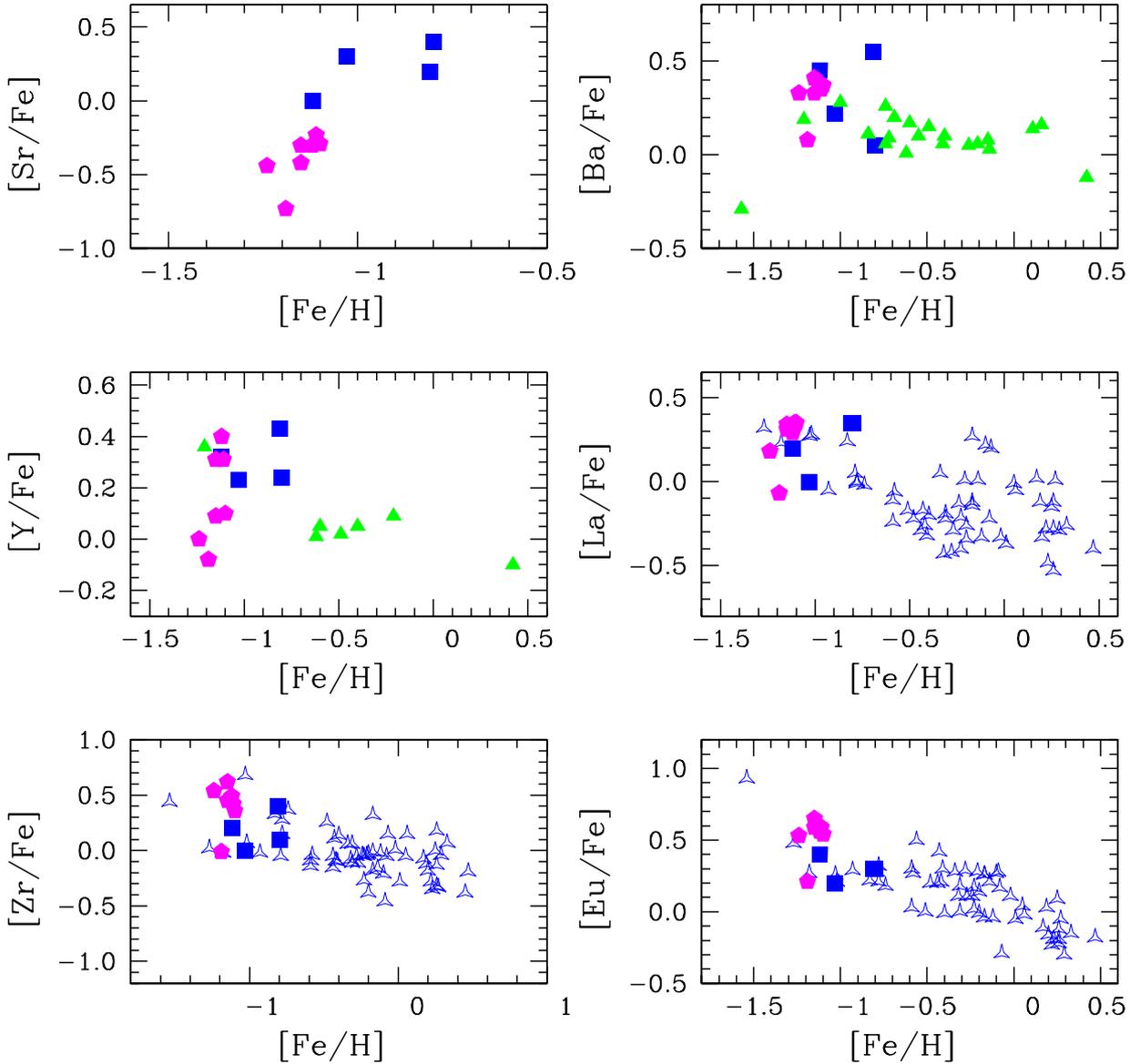,angle=0.,width=17.0 cm}
\caption{Sr, Y, Zr, La, Ba, and Eu-over-Fe for the four
 sample stars (blue filled squares),
compared with literature abundances from
 Johnson et al. (2012) (open triangles), Yong et al.
2014 (magenta filled pentagons), and
 Bensby et al. (2013) (green filled trianges).}
\label{plotheavy} 
\end{figure*}

In Fig. \ref{plotheavy} we plot our results for the heavy elements 
Sr, Y, Zr, La, Ba, and Eu-over-Fe for the four
 sample stars (blue squares) compared with
literature abundances from Johnson et al. (2012), Yong et al.
(2014), and Bensby et al. (2013).
This figure indicates that Sr, Y, and Zr are enhanced, and show a spread
 at [Fe/H]$\sim$$-$1.0, while
Ba and La are also enhanced around [Fe/H]$\sim$$-$1.0. These features
are compatible with expectations from massive spinstars (see Chiappini 2013).

 In M62, the spread in abundances of Sr, Y, and Zr
is consistent with s-process production by massive spinstars.
On the other hand, the observed [Rb/Y] and partially the [Rb/Zr]
seem to disagree with this scenario. The uncertainties of explosive
 nucleosynthesis in core-collapse supernovae are quite large, 
and in particular, the spread of [Rb/Zr] and [Rb/Y] in the yields 
from the SN explosion of spinstars still needs to be explored.



Our results show  enhancements of 
[O/Fe]=+0.36, [Mg/Fe]=$\approx$+0.23, [Si/Fe]=[Ca/Fe]$\approx$+0.13,    
 [TiII/Fe]$\approx$+0.17, and lower [TiI/Fe]$\approx$+0.04.
The r-process element Eu is enhanced by [Eu/Fe]=+0.30.
 The s-elements La and Ba are enhanced with [La/Fe]=+0.23 and [Ba/Fe]=+0.32,  
the latter not as high as the [Ba/Fe]=+0.49 previously found by B09.
The odd-Z element Na shows essentially a solar ratio with  [Na/Fe]$\sim$+0.05,
and [Al/Fe]=+0.20 with the correction for a trend with effective temperature.
 Therefore there is no clear indication of a Na-O anticorrelation. 

The $\alpha$-element enhancements in  O, Mg together with that of 
the r-process element Eu are indicative of a fast early enrichment by
 supernovae type II.

 
 When the  [Ba/Eu] ratio is considered as indicative of the s-process and
 r-process relative contribution, the La and Ba 
abundances, which are higher than Eu, are not consistent with a pure r-process
 contribution ([Ba/Eu]$\sim$0.0 compared with [Ba/Eu]$_{\rm r}$$\sim$$-$0.8, 
Bisterzo et al. 2014). Therefore, many of the conclusions 
discussed by B09 and C11 are still valid.

Fig. \ref{plotheavy} shows the abundances of Sr, Y, Zr, La, Ba, and Eu-over-Fe
 determined here, compared with other determinations. The data suggest an increase of [Zr/Fe], [Ba/Fe], [La/Fe], and [Eu/Fe] towards low metallicities. For these elements, the abundances measured by different authors for the most metal-poor bulge stars seem to agree with each other (i.e. abundances of field stars from Johnson et al. and Bensby et al., and abundances in the globular clusters NGC 6522 and M62). The situation seems to be different for Sr and Y.
There is still very few measurements in the literature for these elements, 
especially at metallicities below [Fe/H] $\simless$ $-$1. The observations suggest a strong increase in the dispersion of the [Y/Fe] ratios below [Fe/H] $<$ $-$1 (in particular, there is a large scatter in the [Y/Fe] ratios in the M62 globular cluster). For Sr the situation is even more unclear, with no determination in bulge field stars, and only a few measurements in NGC 6522 and M62. However, as discussed in the previous section, our
Sr abundances are based only on one very faint line. This might explain the very discrepant measurements
 we find in NGC 6522 compared with those of M62. Furthermore, Yong et al. assumed a correction for Non-Local Thermodynamic Equilibrium (NLTE) effects
 for their [Sr/Fe] of $+$0.29 dex, using lines in the blue region
(their spectra cover the full wavelength region 3800 - 9000 {\rm \AA}),
  and we did not take NLTE effects into account in our determinations.
 For a similar NLTE correction for the stars in NGC 6522, the discrepancy
 with M62 stars would be less relevant. We used a
different line, which means that the same correction might not apply. 
Nevertheless, while
we cannot discard that the two clusters have different [Sr/Fe]
ratios, this possibility seems unlikely since the other measured
abundances are consistent, and we qualitatively expect 
  NLTE effects on our line at \ion{Sr}{I} 6503 as well. 
It is important to note that lines in the bluer region are prohibitive
in terms of exposure times and expected S/N ratios, because our sample stars
that are faint in the optical
and are located in rather high extinction regions in the bulge.

\begin{figure*}
\centering
\psfig{file=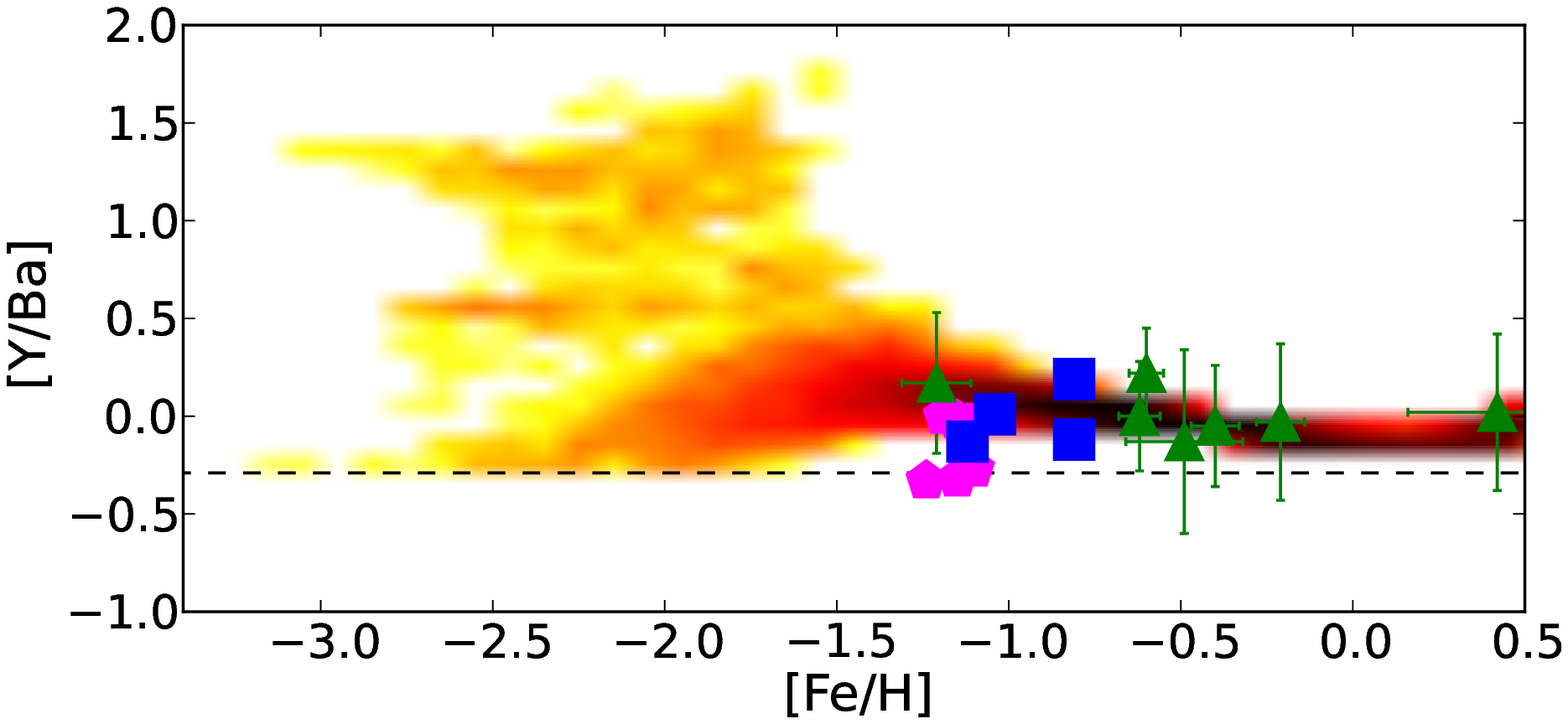,angle=0.,width=13.0 cm}
\psfig{file=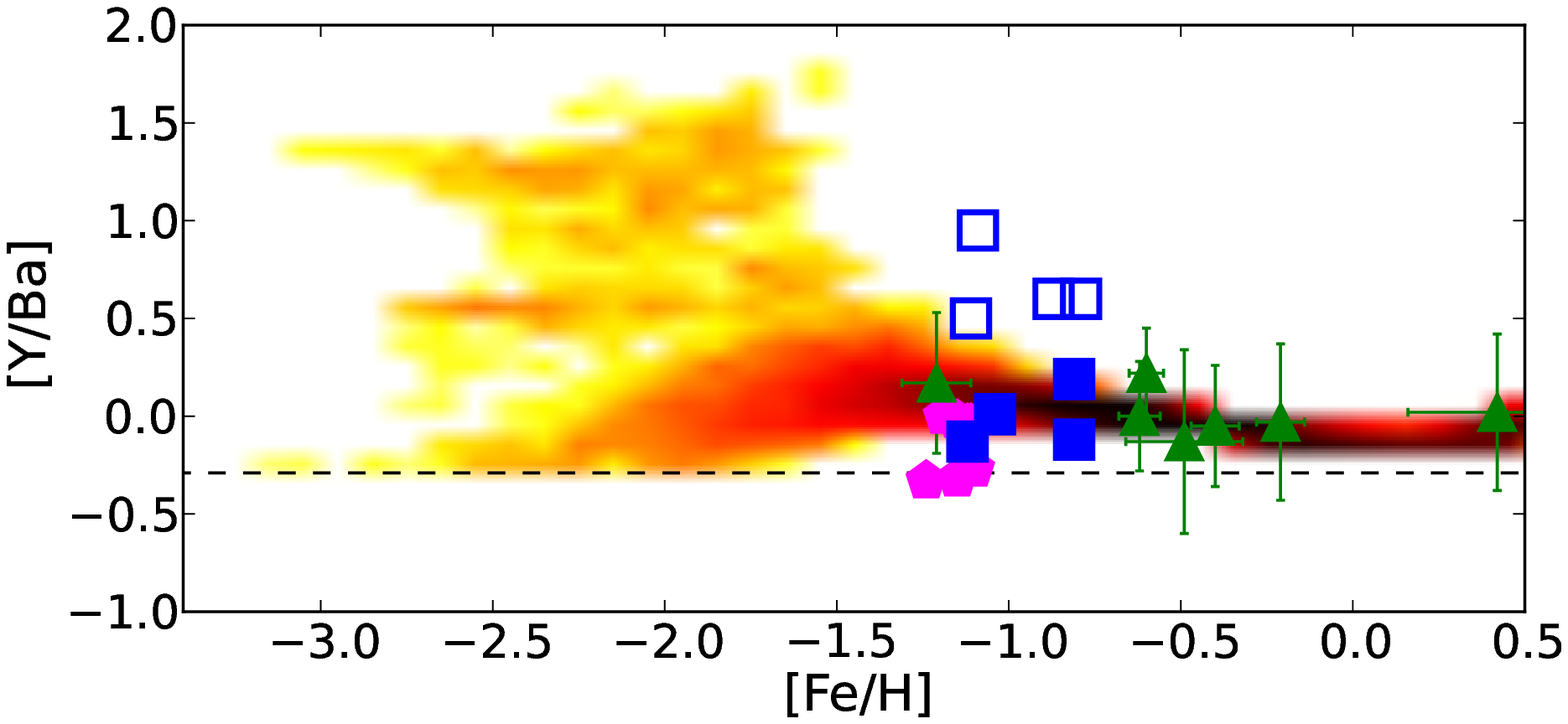,angle=0.,width=13.0 cm}
\psfig{file=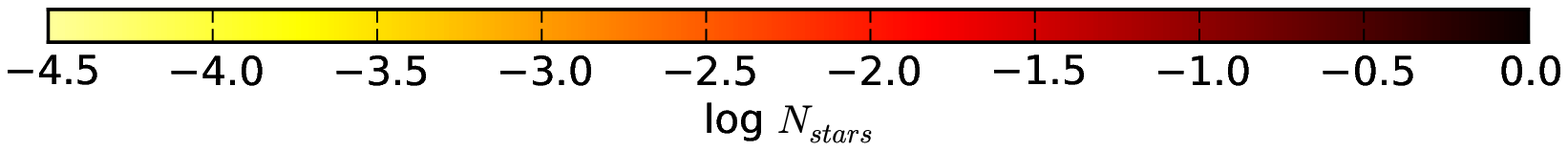,angle=0.,width=13.0 cm}
\caption{Upper panel:
 [Y/Ba] vs. [Fe/H] for the four sample stars (blue filled squares) 
compared with literature abundances from Yong et al.
(2014) for M 62 (magenta filled pentagons) and Bensby et al. (2013) 
for their stars older than 11 Gyr (green filled trianges).
We also show chemical evolution model predictions for the bulge 
(see density scale in the colour bar at the bottom of the figure), 
where the contribution of both spinstars and magneto-rotationally driven 
(MRD) supernovae are taken into account (see Chiappini et al. 2014, 
Cescutti \& Chiappini 2014, Cescutti et al. 2013). 
Lower panel: same as the upper panel, but now also showing the old 
abundance values (open blue squares) for the stars for which
 we have 
re-obtained better chemical abundances (filled blue squares). 
 The dashed line in both panels corresponds to  the r-process fractions, 
which were adopted from Sneden et al. (2008) 
for metal-poor r-element-rich stars.
 Note that in all panels we have 
excluded star B-108.}
\label{plotratios} 
\end{figure*}

 Fig. \ref{plotratios} shows the [Y/Ba] vs. [Fe/H] diagram. We first focus on the upper panel, where the data plotted are the same sample stars 
as shown in Fig. 15 (our new four measurements in NGC 6522 and the literature data).  We also show in this figure a chemical evolution model for the
 Galactic bulge where the enrichment in heavy elements takes place both in spinstars and in magneto-rotationally driven (MRD) supernovae (Chiappini et al. 2014, in prep.; see also Cescutti \& Chiappini 2014; Cescutti et al. 2013). As explained in the introduction, at the time C11 was published there were still no available grids of very metal-poor stellar models of fast-rotating spinstars.
 The situation has now improved 
 and we can now compute chemical evolution models that include
 the contribution of spinstars also for heavy elements such as Y and Ba. From
 our most recent model predictions we find that
 although for the halo the contribution of spinstars leads to a large scatter in the [Y/Ba] ratios around [Fe/H] = $-$3 (Cescutti et al. 2013, Cescutti \& Chiappini 2014), for the bulge the scatter is seen at a lower metallicity around
$-$2 $<$ [Fe/H] $<$  $-$1. This is particularly true in the case shown here,
 where the site for the r-process
 is assumed to be MRD supernovae). The main reason for this is the star formation rate in the bulge that is faster than in the halo. 
More details are reported in Chiappini et al. (2014, in prep.). In the lower panel we also show our old measurements for the re-observed stars. It is clear from this figure that the new chemical abundances obtained here are agree
 better with our model predictions for the bulge. We recall that a good agreement with the [Y/Ba] ratio observed in halo stars is also obtained for a halo model adopting the same nucleosynthesis prescriptions as those
 adopted in the bulge model shown here (Cescutti \& Chiappini 2014).
The star B-108 was discarded, as discussed in Sect. 4.2.

As discussed in C11, the s-element excesses might
 be due to an s-process enrichment
of the primordial matter from which the cluster formed,  or to the s-process
 that occurred in nearby AGB stars
during He-shell flash episodes, and the ejected material would then have been
accreted by the sample stars during their formation process. This
latter explanation would also account for some spread in the 
Ba abundances, hence offering an alternative explanation to the one involving the existence of spinstars in the early chemical enrichment of the Universe. 
C11 have shown that for five out of eight studied stars in NGC 6522 both scenarios would work,
 while for the remaining three stars (with the highest [Y/Ba] ratios) 
the abundances were not compatible with s-process nucleosynthesis in AGB stars,
 and this indicated 
that for these the early enrichment due to spinstars might
 be an option that would have to be investigated by the computation of
 chemical evolution models when stellar yields for spinstars would be available. Our most recent chemical evolution models
 for the bulge that include the contribution of spinstars suggest that the previous large enhancements reported in these three stars are
 incompatible with new calculations for the spinstar scenario. 
Among these three stars only one (B-130) was re-observed,
 and its new abundance agrees well with our model predictions. 
Given this situation, it would be important to try to secure high quality data
 for the other star F-121 as well. 
The interpretation that the  observed enhancements in La, Ba, Y and Sr
 in these very old bulge stars are due to early massive spinstars (C11) is still valid, but not anymore unique.

It is important to stress that the suggestions made in C11 
opened a new field of evidences 
concerning bulge stars. Fig. \ref{heavy} shows that at the metallicity
around [Fe/H]$\sim$$-$1.0, Johnson et al. (2011),
Bensby et al. (2013), and Yong et al. (2014) together with our revised
results indicate that the s-elements [Y/Fe], [Zr/Fe],
[Ba/Fe] and [La/Fe] do show significant enhancements. [Sr/Fe] is moderately
enhanced in the present stars and deficient in M62 (Yong et al. 2014),
 but the abundances for this element are very uncertain.
The enhancement of the r-element Eu is expected, and confirms that
its long-established behaviour is similar to that of $\alpha$-elements.

\section{Conclusions}

We present new abundances for four stars in the oldest known 
MW globular cluster, NGC 6522.
The new abundances were obtained from high-resolution
UVES spectra, with a high signal-to-noise. 
These stars had been studied with lower resolution 
spectra by Barbuy et al. (2009) and Chiappini et al. (2011).  
A fifth star, B-108, was found to be displaced relative to the 
central J2000 coordinates in the OGLE-II catalogue (Udalski et al. 2002), 
which we verified by comparing them with ACS/HST images, 
and therefore  caution in future studies on 
bulge stars is recommended. 
Furthermore, the image shows that there is a blend of two stars, 
with similar radial velocities. This is confirmed by the asymmetries 
observed in the new high-resolution UVES spectra.
B-108 was finally discarded from the sample because of a clear 
contamination of its spectrum, and previous abundances 
for this star cannot be considered either. It would be useful to observe both
 of these stars and examine whether the abundance 
enhancements detected in C11 are present.
It would be very interesting to obtain high-resolution spectra of 
 star F-121, for which an extremely high [Y/Ba] ratio had been estimated 
from FLAMES-GIRAFFE spectra in C11, in which only one better stronger line
is blended, and other lines are faint.
 Nevertheless, we would expect that then the new Y 
measurement will be lower in agreement with the results presented in this work.

The four re-analysed stars now show abundances that are
 compatible with other recent 
measurements of heavy elements (Ba, Y, La, Zr, and Eu) in the Galactic  bulge 
(Yong et al. 2014; Bensby et al. 2013; Johnson et al. 2014). 
An enhancement in s-process-dominant elements is found. In particular, 
the [Y/Ba] ratio of bulge stars in the field (Bensby et al. 2013)
 or in bulge metal poor globular clusters such as NGC 6522 (this work)
  and M62 (Yong et al. 2014) are very similar, and 
lower than indicated by the FLAMES-GIRAFFE spectra analysed in B09 and C11.

The new abundances agree much better 
agreement with recent inhomogenous chemical evolution models where 
spinstars were taken into account. Indeed, the theoretical predictions,
which are now possible to compute thanks to the new grid of spinstar 
models published by Frischknecht et al. (2012), indicate a moderate 
enhancements of the [Y/Ba] ratio at metallicities [Fe/H]$\sim$-1.0.
 The milder enrichment found for Y compared with Ba, and the
 abundance scatter found for Ba and La makes the mass
 transfer contribution from s-process-rich AGB stars
 another possible scenario to explain the signature in NGC6522.
The observation of Pb abundance, if feasible, would help to
 distinguish between these two scenarios, since AGB
 stars at [Fe/H]$\sim$-1.0 are also able to efficiently produce Pb.



\begin{acknowledgements}
We are grateful to the referee for pertinent and useful comments.
BB, EC, and MT acknowledge grants and fellowships from CNPq, Capes and Fapesp.
MT acknowledges the FAPESP post-doctoral fellowship no. 2012/05142-5.
MP acknowledges the support from the Ambizione grant of the SNSF (Switzerland).
DM acknowledges support from the  BASAL Center for Astrophysics and Associated
 Technologies PFB-06 and FONDECYT Project 1130196.
 MZ acknowledges FONDECYT Project 1110393.
DM and MZ also acknowledge support from the Millennium Institute 
of Astrophysics MAS IC-12009.
SO acknowledges the Italian Ministero dell'Universit\`a e della Ricerca
Scientifica e Tecnologica (MURST), Italy. 
\end{acknowledgements}


\begin{appendix}

\section{Equivalent widths and Atomic data}

In Table \ref{balines1} we list the hyperfine structure constants
for the BaII 6141.713 and 6496.897 {\rm \AA} lines, in Table \ref{hfsBa}.2
 the list of lines substructured because of the hyperfine
structure.
In Table \ref{lalines1} we list the hyperfine structure constants for
LaII lines studied in this work.

In Table \ref{srblends} we report the lines that blend and overlap
 \ion{Sr}{I} 6503.989, 6550.244 {\rm \AA}, 
and \ion{Y}{II} 6613.733.

\begin{table*}
\begin{flushleft}
\caption{Atomic constants for BaII used to compute the hyperfine structure:
A constants are taken from Rutten (1978), B constants from Biehl (1976), and 
when the B constants were not available in the literature we assumed them
to be null.
}             
\label{balines1}      
\centering          
\begin{tabular}{lc@{}c@{}c@{}c@{}c@{}c@{}c@{}c@{}c@{}c@{}c@{}c@{}c} 
\noalign{\smallskip}
\hline\hline    
\noalign{\smallskip}
\noalign{\vskip 0.1cm} 
species & $\lambda$ ({\rm \AA}) & \phantom{-}Lower level 
& \phantom{-}J &\phantom{-}A(mK)& \phantom{-}A(MHz) 
&\phantom{-}B(mK)& \phantom{-}B(MHz) & \phantom{-}Upper level 
& \phantom{-}J &\phantom{-}A(mK)& 
\phantom{-}A(MHz) &\phantom{-}B(mK)& \phantom{-}B(MHz)  \\
\noalign{\vskip 0.1cm}
\noalign{\hrule\vskip 0.1cm}
\noalign{\vskip 0.1cm}
$^{135}$BaII & 6141.713 & 5d $^2$D$_{5/2}$ &  5/2 &1.49& 44.6691 &0& 0 & 6p $^2$P$^{\circ}$$_{3/2}$ & 3/2 &+3.47& 104.028 &+2.2& 65.9544  \\
$^{137}$BaII & 6141.713 & 5d $^2$D$_{5/2}$ &  5/2 &1.66& 49.7655 &0& 0 & 6p $^2$P$^{\circ}$$_{3/2}$ & 3/2 &+3.88& 116.3195 &+3.25& 97.4326  \\
$^{135}$BaII & 6496.897 & 5d $^2$D$_{3/2}$ &  3/2 &3.56& 106.7261 &0& 0 & 6p  $^2$P$^{\circ}$$_{1/2}$ & 1/2 &+21.7 & 650.5497 &0& 0  \\
$^{137}$BaII & 6496.897 & 5d $^2$D$_{3/2}$ &  3/2 &3.97& 119.0176 &0& 0 & 6p  $^2$P$^{\circ}$$_{1/2}$ & 1/2 &+24.2& 725.4979 &0& 0  \\
\noalign{\vskip 0.1cm}
\noalign{\hrule\vskip 0.1cm}
\noalign{\vskip 0.1cm}  
\hline                  
\end{tabular}
\end{flushleft}
\end{table*}

\begin{table*}
\begin{flushleft}
\label{hfsBa}
\caption{Hyperfine structure for \ion{Ba}{II} lines.
log gf sources: Kur\'ucz webpage;
NIST; VALD; adopted: fitting on solar and Arcturus spectra.}
\centering
\begin{tabular}{ccccccccc}
\hline
\noalign{\smallskip}
\cline{1-3} \cline{5-7} \\ 
 \multicolumn{3}{c}{6141.713$\rm \AA$; $\chi$=0.7036 eV}  &&
\multicolumn{3}{c}{6496.897$\rm \AA$; $\chi$=0.6043 eV}  & \\
\multicolumn{3}{c}{log gf(total) =  \phantom{+}0.0}  && 
\multicolumn{3}{c}{log gf(total) = $-$0.32}  & \\
\noalign{\smallskip}
\cline{1-3} \cline{5-7} \\
$\lambda$ ({\rm \AA}) & log gf & iso && $\lambda$ ({\rm \AA}) & log gf &iso\\
\noalign{\smallskip}
\cline{1-3} \cline{5-7} \\
 6141.713 & $-$1.617 & 134 &&  6496.897 & $-$1.937 & 134 & \\
 6141.709 & $-$2.385 & 135 &&  6496.903 &$-$2.704 & 135 & \\
 6141.709 & $-$2.431  &135 &&  6496.904 &$-$2.306 & 135 & \\ 
 6141.708 & $-$3.385 &135 && 6496.886 & $-$3.005 & 135 & \\
 6141.712 & $-$2.063  &135 && 6496.907 &$-$2.306 & 135 &  \\
 6141.710 & $-$2.318 &135 && 6496.889 &$-$2.306 & 135 &    \\
 6141.705 & $-$3.561 &135 && 6496.894 & $-$1.858 & 135 &  \\ 
 6141.714 & $-$1.813 &135 && 6496.897 & $-$1.425 & 136 &   \\
 6141.709 & $-$2.415 &135 && 6496.903 &$-$2.475 & 137 &  \\   
 6141.715 & $-$1.607 & 135 && 6496.905 &$-$2.077 & 137 &  \\
 6141.713 & $-$1.105 & 136 && 6496.885 &$-$2.776 & 137 &  \\ 
 6141.713 & $-$2.154 &137 &&   6496.909 &$-$2.077 & 137 &  \\
 6141.713 & $-$2.200 &137 && 6496.888 &$-$2.077 & 137 &   \\         
 6141.712 & $-$3.154 & 137 &&   6496.893 & $-$1.630 & 137 & \\
 6141.714 & $-$1.832 & 137 &&   6496.897 & $-$0.464 & 138   & \\       
 6141.713 & $-$2.087 & 137 &&        &         &     & \\        
 6141.707 & $-$3.330 & 137 &&        &         &     & \\      
 6141.715 & $-$1.582 & 137 &&     &         &     & \\     
 6141.709 & $-$2.184 & 137 &&        &         &     & \\    
 6141.711 & $-$1.376  & 137  &&         &         &     & \\ 
 6141.713 & $-$0.144  & 138  &&         &         &     & \\     
\noalign{\smallskip}
\hline
\end{tabular}
\end{flushleft}
\end{table*}

\begin{table*}
\begin{flushleft}
\caption{Atomic constants for LaII used to compute hyperfine structure:
A and B constants are from Lawler et al. (2001a) and Biehl (1976), 
and where the B constants were not available in the literature,
 we assumed them as null.
}             
\label{lalines1}      
\centering
\begin{tabular}{lc@{}c@{}c@{}c@{}c@{}c@{}c@{}c@{}c@{}c@{}c@{}c@{}c} 
\noalign{\smallskip}
\hline\hline    
\noalign{\smallskip}
\noalign{\vskip 0.1cm} 
species & $\lambda$ ({\rm \AA}) & \phantom{-}Lower level 
& \phantom{-}J &\phantom{-}A(mK)& \phantom{-}A(MHz) 
&\phantom{-}B(mK)& \phantom{-}B(MHz) & \phantom{-}Upper level 
& \phantom{-}J &\phantom{-}A(mK)& 
\phantom{-}A(MHz) &\phantom{-}B(mK)& \phantom{-}B(MHz)  \\
\noalign{\vskip 0.1cm}
\noalign{\hrule\vskip 0.1cm}
\noalign{\vskip 0.1cm}
$^{139}$LaII& 6262.287 &  d6s a3D &  4.0 &--- &---  &--- &---  & d4f y3F & 3.0 &4.959 &148.67  &0.15 & 4.5  \\
$^{139}$LaII & 6320.376  &  d6s a1D & 2.0 &31.649 &948.81  &1.66 &49.77  & d4f y3F & 2.0 &12.204 &365.87  &$-$0.13& -3.90  \\
$^{139}$LaII & 6390.477  &  d6s a3D & 2.0 &-37.6 &-1127.22  &1.66 &49.77  & d4f y3F & 3.0 &4.959 &148.67  &0.15 & 4.5  \\
\hline
\noalign{\vskip 0.1cm}
\hline                  
\end{tabular}
\end{flushleft}
\end{table*}  

\begin{table*}
\begin{flushleft}
\caption{Blends near and overlapping the
 \ion{Sr}{I} 6503.989, 6550.244 {\rm \AA}, 
and \ion{Y}{II} 6613.733 lines.}             
\label{srblends}      
\centering          
\begin{tabular}{lccccccccccccc}     
\noalign{\smallskip}
\hline\hline    
\noalign{\smallskip}
\noalign{\vskip 0.1cm} 
species & {\rm $\lambda$} ({\rm \AA}) & {\rm $\chi_{ex}$ (eV)} & {\rm gf$_{Kurucz}$} &
 {\rm gf$_{NIST}$} & {\rm gf$_{VALD}$} & {\rm gf$_{adopted}$}  \\
\noalign{\vskip 0.1cm}
\noalign{\hrule\vskip 0.1cm}
\noalign{\vskip 0.1cm}
 VI & 6503.757 & 3.567  & $-$4.641 & --- & $-$4.641 &  $-$4.641 & \\
Fe1 & 6503.796 & 5.345  & --- & --- &   $-$8.813 & $-$8.813 & \\
Sc2 & 6503.880 & 7.439  & $-$2.240 & ---  & $-$2.240 & $-$2.240 & \\  
Tm2 & 6503.930 & 3.715  & --- & --- &   $-$2.490 &  $-$2.490 & \\  
Fe1 & 6503.974 & 5.538  & --- & --- &   $-$4.97 & $-$4.970 & \\  
Ce1 & 6503.977 & 0.493  & $-$1.661 & --- &   $-$1.661 & $-$1.661  \\ 
Sr1 & 6503.989 & 2.258  & +0.260 & $-$0.050 &  +0.320   & +0.320 \\ 
Ni1 & 6504.063 & 5.363  & $-$2.215 & --- &  $-$2.415 & $-$2.415 & \\
Pr1 & 6504.066 & 1.291  & $-$0.217 & --- &   $-$0.217 & $-$0.217 & \\  
Ce1 & 6504.125 & 0.517  & $-$1.637 & --- &    $-$1.637 & $-$1.637 & \\ 
 V1 & 6504.164 & 1.183  & $-$1.230 & $-$1.230 &   $-$1.23 & $-$1.230 & \\  
Fe1 & 6504.182 & 4.733  & $-$0.357 & --- &   $-$3.425 &  $-$3.425 & \\ 
Co1 & 6504.210 & 3.687  & $-$1.737 & --- & $-$1.737   & $-$2.437 & \\ 
\noalign{\vskip 0.1cm}
\noalign{\hrule\vskip 0.1cm}
\noalign{\vskip 0.1cm}
Mn2 & 6550.103 & 4.779 & $-$6.804  & --- & $-$6.804 &  $-$6.804     &  \\
Cr1 & 6550.135 & 3.369 & $-$4.185  & --- & $-$4.185 &  $-$4.185     &  \\
Nd2 & 6550.178 & 0.321 &$-$-2.280  & --- & $-$1.850 &  $-$1.850    &  \\
Ca1 & 6550.217 & 5.049 & $-$1.007 & --- &$-$2.107 &  $-$2.107    &  \\
Fe2 & 6550.227 & 4.837 & ---  & --- & $-$6.086 & $-$6.086     &  \\
Fe2 & 6550.249 & 9.761 & ---  & --- & $-$4.468 &  $-$4.468     &  \\
Sc2 & 6550.253 & 7.482 & +0.611  & --- &  +0.611 &  +0.611    &  \\
Ni1 & 6550.355 & 5.363 & $-$1.417 & --- &$-$2.117 &  $-$2.117    &  \\
Si1 & 6550.365 & 6.083 & ---  & --- &$-$2.852 &  $-$2.852    &  \\
SrI & 6550.244 & 2.690  & +0.180   & +0.460  &  +0.180 & +0.180  \\ 
\noalign{\vskip 0.1cm}
\noalign{\hrule\vskip 0.1cm}
\noalign{\vskip 0.1cm}
Ce2 & 6613.392 & 4.014 & ---     & --- & $-$3.090 & $-$1.590 & \\
Fe1 & 6613.502 & 5.538 & ---     & --- & $-$4.412 & $-$3.412 & \\
Ti1 & 6613.599 & 3.718 & ---     & --- & $-$2.025 &  $-$2.325 & \\ 
Ti1 & 6613.620 & 2.495 & ---     & --- &  $-$2.443&  $-$2.743  & \\ 
Ti1 & 6613.626 & 1.460 & $-$2.736  & --- &  $-$3.101&  $-$3.401 & \\
 V2 & 6613.632 & 9.342 & ---     & --- & $-$2.796 & $-$2.796 & \\
Fe2 & 6613.709 & 12.199 & ---    & --- & $-$5.695  & $-$5.695 & \\
 Y2 & 6613.730 & 1.748 & $-$1.110  & $-$1.110 &  $-$1.110 &  $-$1.200 & \\
Cr2 & 6613.776 & 11.234&  ---    & --- &  +0.515 &   +0.515 & \\
Fe1 & 6613.825 & 1.011 & $-$6.689  & --- &  $-$5.587&  $-$5.800 & \\
Ca1 & 6613.899 & 5.882 & $-$3.654  & --- &  $-$3.654&  $-$2.654 & \\
\noalign{\vskip 0.1cm}
\noalign{\hrule\vskip 0.1cm}
\noalign{\vskip 0.1cm}  
\hline                  
\end{tabular}
\end{flushleft}
\end{table*}

\begin{figure*}
\centering
\psfig{file=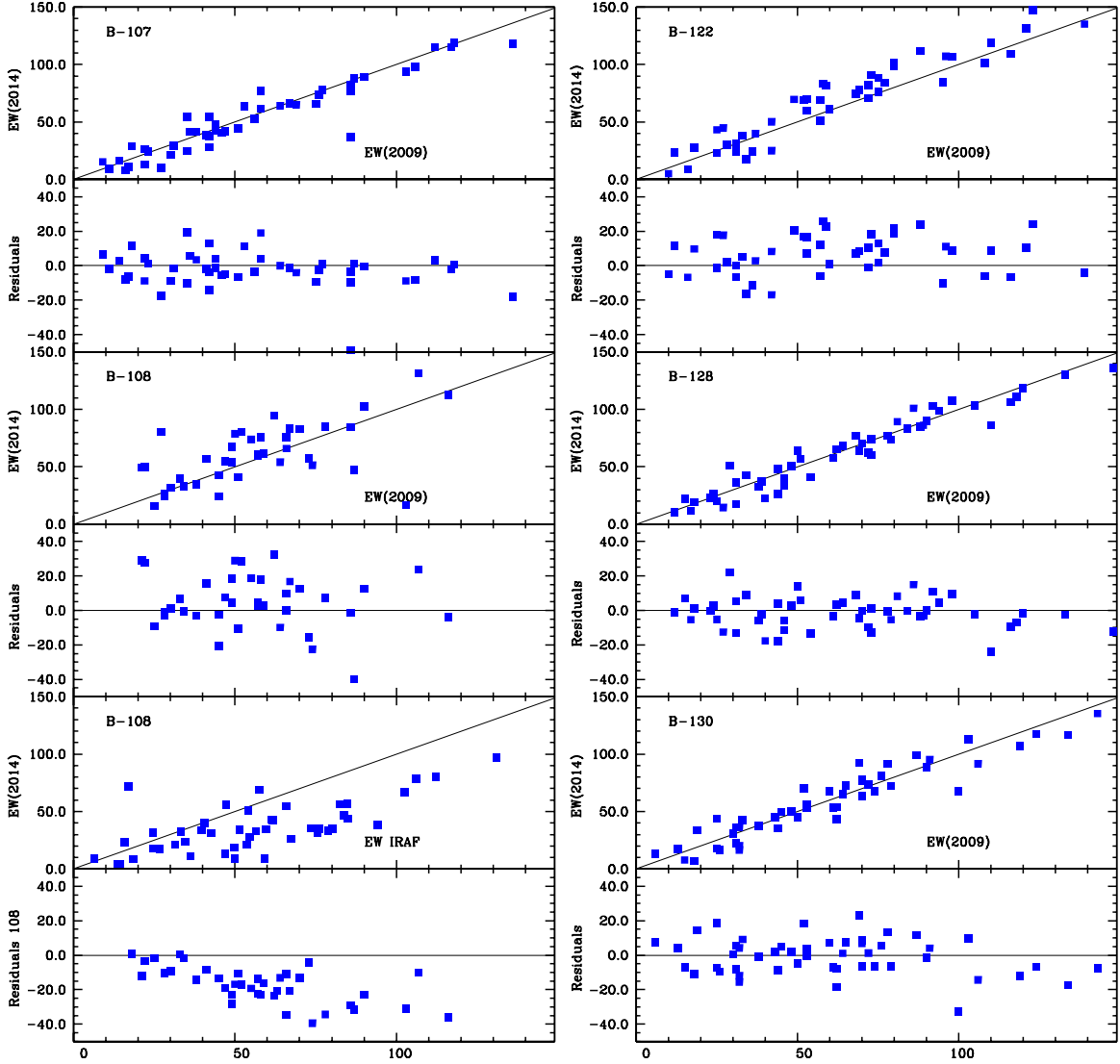,angle=0.,width=17.0 cm}
\caption{Equivalent widths measured in the UVES spectra in 2014 (EW(2014))
vs. equivalent widths measured in the GIRAFFE spectra in 2009 (EW(2009)),
and residuals between the two. For each star the upper panel gives
EW(2014) vs. EW(2009), and the lower panel gives the residuals.
For star B-108, also the measurements with IRAF on the UVES spectra are
 compared with EW(2009).}
\label{residuals} 
\end{figure*}

Table \ref{EW} presents the equivalent widths of
FeI and FeII lines used in the analysis.
In Fig. \ref{residuals} we plot the 
EWs measured in the UVES spectra in 2014 (EW(2014))
vs. EWs measured in the GIRAFFE spectra in 2009 (EW(2009)),
and residuals between the two.
For star B-108, we also compare the measurements 
with IRAF on the UVES spectra are
 with EW(2009). The better de-blending of lines with IRAF from the
probable presence of two stars explains the poor fit between IRAF and 
the other EWs.

\begin{table*}
\label{EW}
\caption[10]{\ion{Fe}{I} and \ion{Fe}{II} lines employed, their wavelengths,
excitation potential (eV), oscillator strengths adopted from B09, and
 equivalent widths as given in B09 from GIRAFFE spectra, and those we
 measured with the code DAOSPEC on UVES spectra. For star B-108 we also measured
EWs with IRAF.}
\begin{flushleft}
\begin{tabular}{lrrrrrrrrrrrrrrrrr}
\noalign{\smallskip}
\hline
\noalign{\smallskip} 
\hbox{\rm species} & ${\rm \lambda}$({\rm \AA}) & $\chi_{\rm ex}$ & \hbox{log~gf}
& \multicolumn{2}{c}{ B-107} & \multicolumn{3}{c}{ B-108} &
\multicolumn{2}{c}{ B-122} &
\multicolumn{2}{c}{ B-128} & \multicolumn{2}{c}{ B-130}  \\
\noalign{\smallskip} 
\hline
\noalign{\smallskip}
&\hbox{\rm \AA} & (eV) & & B09 & DAO & B09 & DAO & IRAF & B09 & DAO & B09 & DAO & B09 & DAO  \\ 
\noalign{\smallskip} 
\hline
\noalign{\smallskip}
FeII &6149.26 & 3.89 &$-$2.69 & 31  &       29.6 &    22   & 49.9    & 18.6  & 25     &  43.1 &  31  &  36.1  & 32 &   36.3  \\
FeII &6247.56 & 3.89 &$-$1.98 & 51  &       44.3 &    25   & 15.7    & 23.4  & 42     &  25.1 &  44  &  26.2  & 43 &  ---    \\
FeII &6432.68 & 2.89 &$-$3.57 & 42  &       38.0 &    28   & 24.7    & 17.7  & 33     &  38.2 &  38  &  32.3 & 38 &   37.2  \\
FeII &6456.39 & 3.90 &$-$2.05 & 76  &       73.6 &    49   & 53.6    & 20.9  & 53     &  60.2 &  51  &  57.0  & 53 &  56.4    \\
FeII &6516.08 & 2.89 &$-$3.31 & 53  &       64.2 &    34   & 33.3    & 32.4 & 44      &  --- &  44  &  47.7  & 43 &   44.9 \\
FeI &6151.62 & 2.18 &$-$3.299 &  44 &      42.9 &     52 &     80.3  & 35.0 &      59 &      81.8  &     68 &       77.1  &     52 &     70.1 \\ 
FeI &6159.38 & 4.61 &$-$1.97 &--- &      11.6 &   --- &    ---   &  2.4 &    --- &     ---   &   --- &       18.8  &     13 &    ---   \\
FeI &6165.36 & 4.14 &$-$1.549 &  23 &      24.1 &   --- &     59.1  &  9.0 &  --- &      54.0  &     29 &       51.1  &     25 &     43.8  \\
FeI &6173.34 & 2.22 &$-$2.879 &   67 &      65.8 &     62 &     94.2  &   38.2 &    88 &     111.9  &     92 &      103.1  &     78 &     91.3  \\
FeI &6180.20 & 2.73 &$-$2.784 &  44 &      42.7 &     50 &     78.8  &  33.1 &     58 &      83.6  &     70 &       70.0  &     60 &     67.3  \\
FeI &6187.99 & 3.94 &$-$1.718 &  22 &      26.1 &     21 &     49.8  &  9.0 &    27 &      44.7  &     34 &       43.1  &     31 &     36.6  \\
FeI &6200.31 & 2.61 &$-$2.437 &  64 &      64.0 &     66 &     75.6  &  31.2 &    73 &      91.3  &     81 &       89.2  &     76 &     81.3  \\
FeI &6213.43 & 2.22 &$-$2.646 &  87 &      88.4 &     78 &     85.0  & 43.7 &     96 &     107.1  &     98 &      107.5  &     91 &     95.1  \\
FeI &6220.78 & 3.88 &$-$2.460 &  11 &       8.9 &   --- &    ---   &   6.0 &  --- &     ---   &   --- &       13.1  &   --- &    --- \\
FeI &6226.74 & 3.88 &$-$2.202 &  16 &       8.2 &   --- &      6.4  & 9.0 &     16 &       9.2  &     17 &       11.6  &     15 &      7.8  \\
FeI &6240.65 & 2.22 &$-$3.388 &  86 &      36.8 &     45 &     24.6  & 31.4 &     57 &      51.1  &     73 &       60.0  &     50 &     45.3  \\
FeI &6246.32 & 3.60 &$-$0.956 &  86 &      82.7 &     73 &     57.6  & 68.8 &       95 &      84.8  &    110 &       86.2  &    100 &     67.4  \\
FeI &6252.56 & 2.40 &$-$1.687 & 117 &     115.0 &    103 &     17.0  &  72.3 &     116 &     109.6  &    118 &      111.1  &    119 &    106.7  \\
FeI &6254.25 & 2.28 &$-$2.480 & 103 &      94.1 &     87 &     47.2  & 55.6 &    108 &     101.8  &    116 &      106.3  &    106 &     91.6  \\
FeI &6270.23 & 2.86 &$-$2.711 &  36 &      41.8 &     51 &     40.7  &  40.4 &    60 &      61.0  &     72 &       62.1  &     62 &     43.4  \\
FeI &6271.28 & 3.32 &$-$2.957 &   9 &      15.3 &     18 &    ---   & 18.6 &   --- &      10.3  &     12 &       10.6  &   --- &      7.4  \\
FeI &6297.79 & 2.22 &$-$2.740 &  58 &      77.0 &     63 &    ---   &  42.6 &    75 &      76.7  &     90 &       89.9  &     72 &     73.2  \\
FeI &6301.50 & 3.65 &$-$0.720 &  90 &      89.5 &     67 &     83.7  &  46.6 &    84 &     ---   &     86 &      100.9  &     90 &     88.9  \\
FeI &6302.50 & 3.69 &$-$0.91 &  75 &      65.6 &     66 &     65.8  & 55.0 &     72 &      82.4  &     88 &       84.5  &     79 &     72.4  \\
FeI &6311.50 & 2.83 &$-$3.224 &  28 &     --- &     38 &     34.7  & 24.0 &      43 &     ---   &     46 &       40.1  &     44 &     35.3  \\
FeI &6315.31 & 4.14 &$-$1.230 &  42 &      27.9 &     33 &     39.7  & 33.5 &     42 &      50.2  &     48 &       50.8  &     33 &     42.1  \\
FeI &6315.81 & 4.14 &$-$1.712 &  30 &      21.2 &     30 &     31.4  &  20.8 &     42 &     ---   &     39 &       37.0  &     30 &     30.5  \\
FeI &6322.69 & 2.59 &$-$2.426 &  69 &      64.9 &     59 &     61.6  &   42.7 &    77 &      84.4  &     89 &       86.0  &     70 &     76.7  \\
FeI &6335.33 & 2.20 &$-$2.229& 106 &      97.7 &     90 &    102.6  & 66.8 &    110 &     118.7  &    120 &      118.3  &    103 &    112.6  \\
FeI &6336.82 & 3.69 &$-$1.053&  77 &      77.9 &     70 &     82.6  &   56.8 &    80 &     102.0  &   --- &       96.7  &     69 &     92.3  \\
FeI &6344.15 & 2.43  &$-$2.922&  56 &      52.4 &     55 &     73.7  &  35.7 &     75 &      88.0  &     84 &       83.7  &     65 &     72.6  \\
FeI &6355.03 & 2.84  &$-$2.40 &  35 &      54.2 &     58 &     76.1  & 34.9 &     80 &      98.4  &     94 &       98.6  &     70 &     78.4  \\
FeI &6392.54 & 2.28  &$-$4.03 &--- &       6.5 &   --- &     36.1  &   10.9 &     18 &      27.7  &     24 &       26.9  &     31 &     22.9  \\
FeI &6393.60 & 2.43  &$-$1.615& 136 &     118.0 &    107 &    131.0  & 97. &     123 &     147.1  &    152 &      143.9  &    143 &    135.2  \\
FeI &6419.95 & 4.73  &$-$0.250&  41 &      38.9 &     47 &     54.5  & 28.1 &     53 &      69.7  &     62 &       65.1  &     53 &     52.8  \\
FeI &6430.85 & 2.18  &$-$2.005& 112 &     115.0 &    116 &    112.3  & 80.2 &     121 &     131.4  &    133 &      130.3  &    124 &    117.3  \\
FeI &6475.62 & 2.56  &$-$2.94 &  42 &      54.7 &     57 &     61.5  & 43.3 &     57 &      68.9  &     64 &       68.4  &     61 &     53.9  \\
FeI &6481.87 & 2.28  &$-$2.984&  58 &      61.7 &     64 &     54.1  & 50.8 &     72 &      71.0  &     78 &       77.2  &     74 &     67.8  \\
FeI &6498.94 & 0.96  &$-$4.699&  44 &      48.0 &     74 &     51.4  &  34.6 &        68 &      74.9  &     73 &       74.1  &     70 &     63.8\\  
FeI &6518.37 & 2.83  &$-$2.748&  46 &      40.5 &     45 &     42.6  &   31.7 &     60 &      61.2  &     61 &       57.9  &     48 &     50.2  \\
FeI &6533.93 & 4.56  &$-$1.453&  19 &     ---  &   --- &    ---   &--- &     25 &     ---   &   --- &      ---   &     25 &    ---   \\
FeI &6569.22 & 4.73  &$-$0.422&  38 &      41.4 &     49 &     67.4  &  26.2 &     52 &      68.9  &     69 &       64.4  &     62 &     54.1  \\
FeI &6574.23 & 0.99  &$-$5.042&  18 &      29.4 &     41 &     56.4  & 32.5 &     49 &      69.6  &     50 &       63.9  &     45 &     50.0  \\
FeI &6575.02 & 2.59  &$-$2.824&  47 &      42.1 &     57 &     59.5  & 34.4 &     69 &      77.5  &     79 &       73.6  &     64 &     65.1  \\
FeI &6591.31 & 4.59  &$-$2.06 &--- &     ---  &   --- &     13.5  &  4.2 &   --- &     ---   &      9 &      ---   &   --- &    ---   \\
FeI &6593.87 & 2.43  &$-$2.422&  86 &      76.4 &     86 &     84.8  &57.1 &      98 &     107.0  &    105 &      103.0  &     87 &     98.9  \\
FeI &6597.56 & 4.80  &$-$1.061&  13 &     --- &     21 &    ---   &--- &     31 &      24.5  &     34 &      ---   &     17 &    ---   \\
FeI &6608.03 & 2.28  &$-$4.038&   7 &     --- &   --- &     18.4  & 8.4 &     25 &      23.5  &     15 &       22.2  &     13 &     17.0  \\
FeI &6678.00 & 2.69  &$-$1.420& 118 &     118.8 &   --- &    106.1  &78.5 &     139 &     135.1  &    148 &      135.6  &    134 &    116.9  \\
FeI &6703.57 & 2.76  &$-$3.15 &  35 &      24.8 &     28 &     26.7  &17.6 &      37 &      39.9  &     54 &       40.7  &     19 &     33.5  \\
FeI &6705.11 & 4.61  & ...  &--- &      17.7 &   --- &     12.4  &--- &      12 &      23.6  &     40 &       22.3  &     26 &     16.6  \\
FeI &6710.32 & 1.48  &$-$4.874&  14 &      16.8 &   --- &     46.9  &  13.5 &    31 &      30.9  &     46 &       34.3  &     32 &     20.1  \\
FeI &6713.75 & 4.80  &$-$1.602&--- &       5.3 &   --- &     14.4  &  4.0 &    --- &     ---   &   --- &        8.1  &      5 &    ---   \\
FeI &6715.38 & 4.59  &$-$1.638&  17 &      10.7 &   --- &     27.1  &--- &      12 &     ---   &     27 &       14.5  &     18 &      6.9  \\
FeI &6716.24 & 4.56  &$-$1.927&  11 &     ---  &   --- &     55.2  & --- &      8 &     ---   &     31 &      ---   &     10 &    ---   \\
FeI &6725.36 & 4.10  &$-$2.300&--- &     ---  &   --- &     32.5  &--- &      10 &       5.0  &      5 &      ---   &     20 &    ---   \\
FeI &6726.67 & 4.59  &$-$1.090&--- &      10.5 &     27 &     80.3  & --- &     34 &      17.9  &     18 &       19.3  &     32 &     16.5  \\
FeI &6733.15 & 4.64  &$-$1.576&  22 &      13.1 &   --- &    ---   &--- &      27 &     ---   &     31 &       17.9  &     18 &      6.9  \\
FeI &6739.52 & 1.56  &$-$4.941&--- &      --- &   --- &     59.1  &--- &      36 &      24.6  &     25 &       19.7  &     25 &     17.7  \\
FeI &6752.71 & 4.64  &$-$1.366&  27 &       9.7 &   --- &     94.2  &--- &      28 &      30.0  &     23 &       22.7  &      6 &     13.5  \\
\noalign{\smallskip} \hline \end{tabular}
\end{flushleft} 
\end{table*}

\end{appendix}

\end{document}